\documentclass[aps,twocolumn,pra,superscriptaddress,amsmath,showpacs,tightenlines]{revtex4}
\usepackage{graphicx}
\usepackage{color}

 \def\fig#1{{#1}}
\def \etal{\textit{et al.}}

\def \Title#1{``#1''}

\newcommand{\jpc}{J. Phys.: Condens. Matter~}

\newcommand{\job}{J. Opt. B: Quantum Semiclass. Opt.~}

\def\tr{{\rm Tr}}
\def\I{{\rm i}}

\def\RE{{\rm Re}}
\def\IM{{\rm Im}}

\def\<{\langle}
\def\>{\rangle}

\newcommand{\ket}[1]{\mbox{$|#1\rangle$}}
\newcommand{\bra}[1]{\mbox{$\langle#1|$}}
\newcommand\fra[2]{{\textstyle{\frac{#1}{#2}}}}
\newcommand\halfpi{{\textstyle{\frac{\pi}{2}}}}

\def \info#1{}

\newcommand\MAT[2]{{\left[\begin{array}{cc} #1 \\ #2 \end{array} \right]}}
\newcommand\MATT[4]{{\left[\begin{array}{cccc} #1 \\ #2  \\ #3  \\ #4 \end{array} \right]}}

\begin{document}


\title{Quantum state tomography
of large nuclear spins in a semiconductor quantum well: Optimal
robustness against errors as quantified by condition numbers}

\author{Adam Miranowicz}
\affiliation{CEMS, RIKEN, 351-0198 Wako-shi, Japan}
\affiliation{Faculty of Physics, Adam Mickiewicz University,
PL-61-614 Pozna\'n, Poland}

\author{\c{S}ahin K. \"Ozdemir}
\affiliation{CEMS, RIKEN, 351-0198 Wako-shi, Japan}
\affiliation{Department of Electrical and Systems Engineering,
Washington University, St.~Louis, Missouri 63130 USA}
\affiliation{Graduate School of Engineering Science, Osaka
University, Toyonaka, Osaka 560-8531, Japan}

\author{Ji\v{r}\'\i~Bajer}
\affiliation{Department of Optics, Palack\'{y} University, 772~00
Olomouc, Czech Republic}

\author{Go~Yusa}
\affiliation{PRESTO-JST, Honmachi, Kawaguchi, 331-0012 Saitama,
Japan}\affiliation{Department of Physics, Tohoku University,
Sendai, 980-8578 Miyagi, Japan}

\author{Nobuyuki Imoto}
\affiliation{Graduate School of Engineering Science, Osaka
University, Toyonaka, Osaka 560-8531, Japan}

\author{Yoshiro Hirayama}
\affiliation{ERATO Nuclear Spin Electronics Project, Aramaki, Aza
Aoba, Sendai 980-0845, Japan} \affiliation{Department of Physics,
Tohoku University, Sendai, 980-8578 Miyagi, Japan}

\author{Franco Nori}
\affiliation{CEMS, RIKEN, 351-0198 Wako-shi, Japan}
\affiliation{Department of Physics, The University of Michigan,
Ann Arbor, Michigan 48109-1040, USA}

\today

\begin{abstract}
We discuss methods of quantum state tomography for solid-state
systems with a large nuclear spin $I=3/2$ in nanometer-scale
semiconductors devices based on a quantum well. Due to quadrupolar
interactions, the Zeeman levels of these nuclear-spin devices
become nonequidistant, forming a controllable four-level quantum
system (known as quartit or ququart). The occupation of these
levels can be selectively and coherently manipulated by
multiphoton transitions using the techniques of nuclear magnetic
resonance (NMR) [Yusa \emph{et al.}, Nature (London) {\bf 434},
101 (2005)]. These methods are based on an unconventional approach
to NMR, where the longitudinal magnetization $M_z$ is directly
measured. This is in contrast to the standard NMR experiments and
tomographic methods, where the transverse magnetization $M_{xy}$
is detected. The robustness against errors in the measured data is
analyzed by using the condition number based on the spectral norm.
We propose several methods with optimized sets of rotations
yielding the highest robustness against errors, as described by
the condition number equal to 1, assuming an ideal experimental
detection. This robustness is only slightly deteriorated, as given
by the condition number equal to 1.05, for a more realistic
``noisy'' $M_z$ detection based on the standard cyclically-ordered
phase sequence (CYCLOPS) method.

\end{abstract}



\maketitle

\section{Introduction}

Quantum state engineering has been attracting increasing attention
in fundamental physics research as well as in applications in
quantum cryptography, quantum communication and, potentially,
quantum information processing (QIP)~\cite{SchleichBook}. Quantum
state engineering provides methods for synthesis of quantum
states, and their coherent control and characterization. The
latter task can be realized by quantum state and process
tomographic methods.

\emph{Quantum state tomography} (QST) is a method for
reconstruction of a quantum state in a series of measurements
performed on an ensemble of identical quantum states.
\emph{Quantum process tomography} (QPT) is a method, closely
related to QST, which enables a complete characterization of the
dynamics of a quantum system. Both QST and QPT have been applied
widely to QIP in finite- and infinite-dimensional optical systems
(for reviews see Refs.~\cite{ParisBook,DAriano03} and references
therein). In particular, much work has been on QST of polarization
states of photons (see, e.g.,
Refs.~\cite{James01,Altepeter05,Burgh08,Adamson10,Miran14}),
homodyne QST~\cite{Vogel89}, and homodyne
QPT~\cite{Lobino08,Wang13} probed with coherent states. Other
examples include QST in superconducting
circuits~\cite{You05,Liu04}. QST has also been applied in
nuclear-spin systems using nuclear magnetic resonance (NMR)
spectroscopy, which was motivated by quantum information
interest~\cite{Jones11,Vandersypen04}. The NMR QST and NMR QPT
were first developed for liquid-state nuclear spin-1/2
systems~\cite{Vandersypen04}, and only later applied both to
liquid- and solid-state systems of quadrupolar nuclei of
spin-3/2~\cite{Bonk04,Kampermann05,Auccaise08,Teles12} and
spin-7/2~\cite{Teles07}.

The use of multi-level systems (so-called \emph{qudits}) instead
of two-level systems (qubits) is an alternative
paradigm~\cite{Nori,Lanyon} in QIP, which has attracted attention
in recent years. Standard examples of higher-order nuclear spins
$I$ include: $I=3/2$ for the isotopes $^{69}$Ga, $^{71}$Ga, and
$^{75}$As (in, e.g., GaAs), $I=5/2$ for $^{27}$Al (in, e.g., AlN)
or $^{121}$Sb (in, e.g., FeSb$_2$), $I=7/2$ for $^{123}$Sb (also
in FeSb$_2$), as well as $I=9/2$ for $^{113}$In and $^{115}$In
(in, e.g., InAs, InSb, and InP), and $^{73}$Ge. Note that large
nuclear spins occur also in molecular magnets, i.e., clusters of
spins, which can be applied for
QIP~\cite{Leuenberger01,Ardavan07}. As another example, the
superconducting circuits in Ref.~\cite{Nori} have up to five
levels and can model rotations of spin-1 and spin-3/2. Here, we
will focus solely on QIP using quadrupolar nuclei with spin-3/2,
which are equivalent to a four-level system.

Another motivation for the application of qudits for QIP is
related to an important question concerning the scalability of two
qubits to many qubits. If one simply plans to increase the number
$N$ of qubits, then the required numbers of levels scales up
exponentially. This becomes very hard to implement when $N$ is
large. However, some ordinary classical computers are not
assembled with simple AND, NAND, OR, NOT gates, but instead they
are constructed using higher-level logic gates. Similarly, quantum
computers might be constructed with slightly more complex logic
gates, rather than, e.g., only single-qubit gates and CNOT gates.
In this direction, multi-level systems are helpful.

A word of caution: Replacing qubits by qudits causes a faster
exponential divergence in the number of levels, thus one loses the
advantage of ``using a single multi-level system'' over ``using
many two-level systems.'' Thus, this approach should be applied
carefully. Indeed, using qudits could reduce the complexity of
quantum computers (see Ref.~\cite{Nori} and the justifications
given there). Moreover, for qudits, there are optimal recipes for
gate operations (see Sec.~III) and quantum tomography methods to
be discussed in the following sections.

Various two-qubit quantum state engineering methods and quantum
algorithms have been realized in NMR experiments with spin-3/2
systems. Examples include: the demonstration of
classical~\cite{Khitrin00,Sinha01,Kumar02} and
quantum~\cite{Sarthour03,Kampermann02,Bonk04,Kampermann05} gates,
generation of Bell states~\cite{Sarthour03,Kampermann02}, the
quantum Fourier transform~\cite{Kampermann05}, and implementations
of simple quantum algorithms (i.e., the two-qubit Grover search
algorithm~\cite{Ermakov02,Kampermann05} and the Deutsch-Jozsa
algorithm~\cite{Das03a,Kampermann05}). The existence of quantum
correlations (as revealed by quantum discord) was also
experimentally demonstrated in spin-3/2 systems (see,
e.g.,~\cite{Soares10}). It is worth noting that a prerequisite for
the realization of all these gates and algorithms is the
preparation of pseudo-pure states (see also
Refs.~\cite{Hirayama06,Jones11,Tan12} and references therein).

Quadrupolar nuclei with spin-7/2 have also attracted increasing
interest, as it is highly desirable to scale QIP beyond two (real
or virtual) qubits. A few NMR experiments were performed with
spin-7/2 systems, e.g., the preparation of effective pure
states~\cite{Khitrin01a}, a quantum simulation~\cite{Khitrin01b},
a half-adder and subtractor operations~\cite{Murali02,Kumar02}, a
test of phase coherence in electromagnetically-induced
transparency~\cite{Murali04}, and three-qubit Deutsch-Jozsa
algorithms~\cite{Das06,Teles07,Gopinath08}.

A complete verification of the generated states and/or
performed algorithms in the aforementioned experiments
requires the application of QST.

In this article, we describe QST methods for an unconventional
approach to NMR (sometimes referred to as ``exotic NMR'') in
semiconductor nanostructures~\cite{Machida03, Yusa05, Hirayama06,
Ota07, decoherence}, which is based on the measurement of the
longitudinal magnetization $M_{z}$.

In contrast to this approach, the vast majority of the NMR
tomographic methods are based on conventional (standard) NMR
experiments, where the transverse magnetization $M_{xy}$ is
detected. Indeed, a very tiny magnetic field produced by the
nuclear spin rotation in the $xy$-plane with a resonant frequency
is picked up by a surrounding coil. In this method, the $M_{xy}$
component is measured by using induction detection ($M_{xy}$
detection). However, this widely used conventional NMR suffers
from low sensitivity arising from induction detection, so one
should prepare large volume samples occasionally reaching a qubic
centimeter (at least a qubic millimeter). In the application to
semiconductor (solid-state) systems (see Ref.~\cite{Hirayama09}
and references therein), multiple-layer quantum wells with 10-100
layers should be prepared to detect clear signals with a
sufficient noise-to-signal ratio. A main advantage of
semiconductor (solid-state) qubits is its precise controllability
by using gate operations. Such gate operation is based on a single
quantum well and nanostructure so conventional NMR is obviously
not appropriate for these systems. Since the mid-2000s, highly
sensitive NMR methods suitable for semiconductor hetero- and
nanosystems have been developed by using
electrical~\cite{Machida03,Yusa05} and optical~\cite{Kondo08}
means. However, they all relied on a direct measurement of the
nuclear spin magnetization, i.e., $M_{z}$ detection. Therefore, it
is important for semiconductor (solid-state) nuclear-spin qubits
to develop QST appropriate for the direct detection of $M_{z}$.

Here we study NMR tomography of solid-state four-level quantum
systems, also known as quartits or ququarts.  The main result of
this paper is the proposal of various QST methods based on $M_z$
detection, which are the most robust against errors as quantified
by a condition number equal (or almost equal) to 1. Note that the
proposed QST methods can be applied not solely to solid-state
systems but also to liquid-state quartits. Moreover, these methods
can be generalized for QST of qudits. There has also been interest
in the generation and state tomography of other systems,
especially optical qudits, including qutrits (i.e., three-level
quantum systems) (see, e.g., Ref.~\cite{Thew02}).


The paper is organized as follows: In Sec.~II, we specify the
quadrupolar interaction model. In Sec.~III, we describe sequences
of NMR pulses for implementing qubit gates in qudits. In Sec.~IV,
we present the key aspects of the $M_z$-based QST of a spin-3/2
system. We also briefly discuss a nanometer-scale all-electrical
resistively-detected NMR device~\cite{Machida03,Yusa05}, where the
$M_z$-magnetization can be measured. In Sec.~V, we discuss the
linear reconstruction of density matrices in relation to condition
numbers describing how these methods are robust against errors. In
Sec.~VI, we specify the $M_z$ detection approaches to be applied
in the next sections. These include three approaches: a
theoretical approach, as well as both ideal and non-ideal (noisy)
experimental approaches. The main results of this paper are
presented in Secs.~VII--IX. Specifically, we propose various sets
of rotations, which enable optimal reconstructions of all the
diagonal (in Sec.~VII) and off-diagonal (in Sec.~VIII) elements of
a spin-3/2 density matrix. These two reconstructions are combined
in Sec.~IX. In Sec.~X, we show how to construct sets of
operationally-optimized rotations by finding single-photon
replacements for multiphoton rotations. We conclude in Sec.~XI. In
the Appendices, for completeness and clarity, we define selective
rotations, and briefly compare the $M_{xy}$ and $M_{z}$
detections.

\begin{figure}

 \fig{ \includegraphics[width=8.2cm]{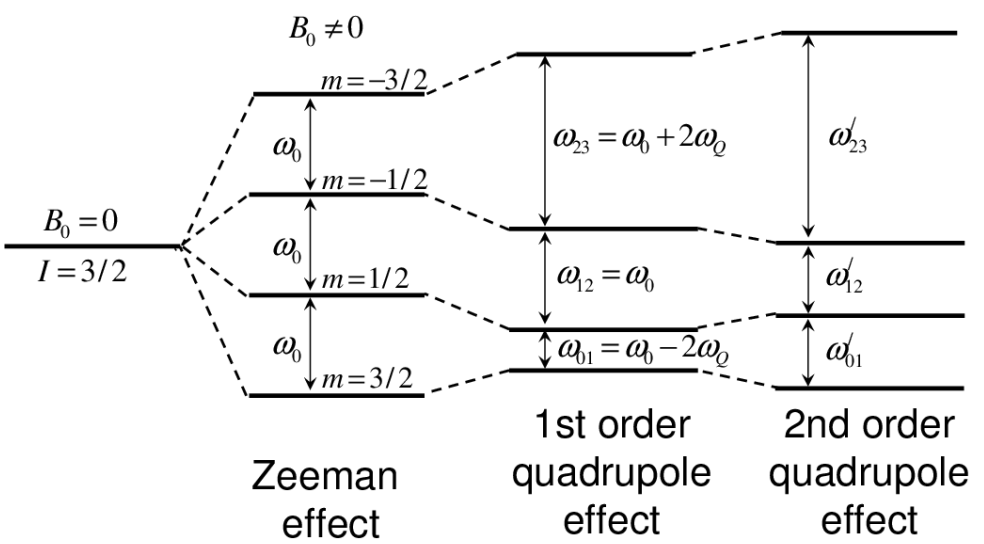}}

 \caption{Schematic diagram of the energy levels of
a spin $I=3/2$ nucleus in an external magnetic field ${B}_0$ due
to Zeeman interactions with additional shifts originating from the
first- and second-order quadrupole interactions. The second-order
shifts completely remove the degeneracy between the levels, which
is desirable for QST. Unfortunately, the second-order shifts are
negligible in the nanoscale device studied here, and thus will be
omitted hereafter.}
\end{figure}
\begin{figure}%

 \fig{ \includegraphics[width=8.2cm]{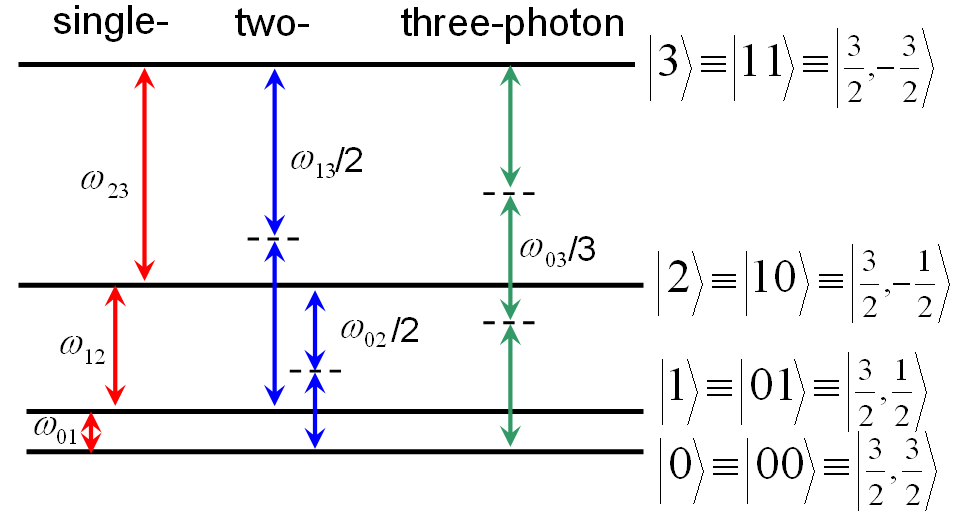}}

\caption{(Color online) A quartit (a spin-3/2 nucleus) is formally
equivalent to two qubits (two spin-1/2 nuclei), as shown here via
their energy levels and their corresponding frequencies.
Single-photon ($\hbar \omega_{01},\hbar \omega_{12},\hbar
\omega_{23}$), two-photon ($\hbar \omega_{02}/2,\hbar
\omega_{13}/2$), and three-photon ($\hbar \omega_{03}/3$)
transitions drive single, double, and triple quantum coherent
oscillations between two levels separated by one ($\Delta m=1$),
two ($\Delta m=2$), and three ($\Delta m=3$) quanta of angular
momentum, respectively. }
\end{figure}

\section{Interaction model}

First, we describe a model for large nuclear spins, in a
semiconductor quantum well, which are interacting with
radio-frequency (RF) pulses. A general description of such an
interaction can be found in standard textbooks on NMR (see, e.g.,
Refs.~\cite{AbragamBook,ErnstBook}). In particular, the model
described in detail by Leuenberger \etal~\cite{Leuenberger02},
which was directly applied in the experiment of Yusa
\etal~\cite{Yusa05}, can also be adapted here.

Specifically, we analyze an ensemble of quadrupolar nuclei (with
spin $I=3/2$) in a semiconductor quantum well interacting with $N$
RF pulses of the carrier frequency $\omega_{_{\rm RF}}^{(k)}$,
phase $\phi_{_{\rm RF}}^{(k)}$, and magnetic-field amplitude
${B}_{k}$ ($k=1,2,...,N$) in the presence of a strong magnetic
field ${B}_{0}$. The effective total Hamiltonian in the laboratory
frame reads~\cite{AbragamBook,ErnstBook,Leuenberger02,Yusa05}
\begin{eqnarray}
 {\cal H}&=&{\cal H}_{0}+{\cal H}_{\rm int},\label{H}
\end{eqnarray}
being a sum of the free term
\begin{eqnarray}
 {\cal H}_{0} &=& {\cal H}_{\rm Z}+{\cal H}_{Q}
  = \hbar\omega_{0}{I}_{z}+\frac{\hbar\omega_{Q}}{3}
  [3{I}_{z}^{2}-I(I+1)], \label{H0}
\end{eqnarray}
and the term describing the interaction of the nuclei with $N$
pulses:
\begin{eqnarray}
  {\cal H}_{\rm int} = \sum_{k=1}^N
  \frac{\hbar\omega_{k}}{2}\left[{I}_{+}{\rm e}^{-i(\omega_{_{\rm RF}}^{(k)}t+\phi_{_{\rm
  RF}}^{(k)})}  + {I}_{-} {\rm e}^{i(\omega_{_{\rm RF}}^{(k)}t+\phi_{_{\rm
  RF}}^{(k)})} \right]\; \\
= \sum_{k=1}^N
  \hbar\omega_{k}\!\left[{I}_{x}\cos(\omega_{_{\rm RF}}^{(k)}t+\phi_{_{\rm
  RF}}^{(k)})  + {I}_{y} \sin(\omega_{_{\rm RF}}^{(k)}t+\phi_{_{\rm
  RF}}^{(k)}) \right]. \label{Hint}
 \nonumber
\end{eqnarray}
Here, ${\cal H}_{\rm Z}=\hbar\omega_{0}{I}_{z}$ and ${\cal H}_{Q}$
describe, respectively, the Zeeman and quadrupole splittings (see
Fig.~1). The operator ${I}_{\alpha}$ (for $\alpha=x,y,z$) is the
$\alpha$-component of the spin angular momentum operator, and $
I_{\pm}= I_x\pm i I_y$. Moreover, $\omega_{0}=-\gamma {B}_{0}$ is
the nuclear Larmor frequency, and $\omega_{k}=-\gamma {B}_{k}$ is
the amplitude (strength) of the $k$th pulse, where $\gamma$ is the
gyromagnetic ratio. For the example of the nuclei ${}^{69}{\rm
Ga}$ and ${}^{71}{\rm As}$ of spin $I=3/2$ in semiconductor GaAs,
we can choose the gyromagnetic ratios to be $\gamma({}^{69}{\rm
Ga})=1.17 \times 10^7 \,{\rm s}^{-1} {\rm T}^{-1}$ and
$\gamma({}^{71}{\rm As})=7.32 \times 10^6\, {\rm s}^{-1} {\rm
T}^{-1}$, which are estimated from the spectra measured in
Ref.~\cite{Yusa05}. The Hamiltonian ${\cal H}_{Q}={\cal
H}_{Q}^{(1)}+{\cal H}_{Q}^{(2)}+...$ describes the quadrupolar
interaction as a sum of the first- and second-order quadrupolar
terms (as shown in Fig.~1), but also higher-order terms. The
first-order quadrupolar splitting parameter (quadrupolar
frequency) $2\omega_{Q}$ is given for solids by~\cite{LevittBook}:
\begin{equation}
\omega_{Q}\equiv \omega_{Q}^{(1)}=\frac{3\pi C_{Q}}{4I(2I-1)}
(3\cos^{2}\theta_{Q}-1),
\end{equation}
where $C_{Q}$ is the quadrupolar coupling constant, and
$\theta_{Q}$ is the angle between the direction of the field
${B}_{0}$ and the principle axis of the electric-field gradient
tensor. We assume a uniaxial electric-field gradient tensor, i.e.,
the biaxiality parameter is zero ($\eta_{Q}=0$). Under the secular
approximation, which is valid for relatively small $\omega_{Q}$,
the effective interaction is described solely by the first-order
quadrupolar Hamiltonian, as we have assumed in Eq.~(\ref{H0}).

The quadrupolar frequencies are typically of the order of
10--100~kHz. For example, the values for the isotopes in
semiconductor GaAs can be found in
Refs.~\cite{Salis01,Leuenberger02,Yusa05}. In our numerical
simulations, we set the following values of the quadrupolar
frequencies $\omega_{Q}({}^{69}{\rm Ga})=15.2$~kHz and
$\omega_{Q}({}^{71}{\rm As})=26.9$~kHz.  These values were
estimated from the experimental spectra reported in
Ref.~\cite{Yusa05}. Moreover, we also choose in our simulations
the same values of parameters as those measured or estimated in
the experiment with nanometer-scale device in Ref.~\cite{Yusa05}.
Namely, ${B}_0=6.3$~T and ${B}_k=\;$0.2--1.4~mT, and decoherence
time is $T_2\approx 1$~ms.

Let us denote the eigenvalues and eigenvectors of ${\cal H}_0$ by
$\epsilon_{m}$ and $|m\rangle$, respectively, i.e., ${\cal
H}_0|m\rangle=\epsilon_{m}|m\rangle.$ If the condition $|\omega_k|
\ll |\omega_Q| \ll |\omega_{0}|$ is satisfied, one can apply a
selective RF pulse resonant with a transition
$|m\rangle\leftrightarrow|n\rangle$, i.e., $\hbar\omega_{_{\rm
RF}}^{(k)} =\epsilon_{m}-\epsilon_{n}$, where $m,n=0,1,...$\,. One
can also analyze $N$-photon resonant transitions, which correspond
to the condition $N\hbar\omega_{_{\rm
RF}}^{(k)}=\epsilon_{m}-\epsilon_{n}$, where $k=1,2,...$ (see
Fig.~2). Note that Eq.~(\ref{Hint}) can still be used, even if the
$\omega_{_{\rm RF}}^{(k)}$ are slightly detuned from the resonant
frequencies by $\delta\omega_k$.

In the more general case when $N$ RF pulses of different
frequencies $\omega_{_{\rm RF}}^{(k)}$ are applied simultaneously,
then clearly the {\em standard} rotating frame is not useful to
transform the time-dependent Hamiltonian, given by
Eq.~(\ref{Hint}), into a time-independent form. However, if the
quadrupolar splitting $2\hbar\omega_Q$ is much larger than the
detuning energies $\hbar\delta\omega_k$, then one can still
transform Eq.~(\ref{H}) into a completely time-independent
Hamiltonian in a {\em generalized} rotating frame, as described
in, e.g., Ref.~\cite{Leuenberger02}.

The Hamiltonian ${\cal H}$ can be transformed to the rotating
frame as follows
\begin{eqnarray}
  {\cal H}_{\rm rot} &=&  U {\cal H}  U^{\dagger}
  -i\hbar  U\frac{\partial  U^{\dagger}}{\partial t}.
\label{HrotU}
\end{eqnarray}
Let us assume that only a single pulse ($k=1$) is applied of
strength $\omega_1$, frequency $\omega_{_{\rm RF}}\equiv
\omega_{_{\rm RF}}^{(1)}$, and phase $\phi\equiv \phi_{_{\rm
RF}}^{(1)}$. Then the time-dependent Hamiltonian, given by
Eq.~(\ref{H}), in the frame rotating with angular frequency
$\omega_{_{\rm RF}}$, becomes the well-known time-independent
Hamiltonian:
\begin{eqnarray}
 {\cal H}_{\rm rot}&=&
 \hbar\Delta\omega{I}_{z} +{\cal
 H}_{Q} +\hbar\omega_{1} {I}_{\phi}\,,\label{Hrot}
\\
 {I}_{\phi}&=&{I}_x\cos\phi+{I}_y\sin\phi
 =\frac12 ({I}_+{\rm e}^{-i\phi}+{I}_-{\rm e}^{i\phi}),
  \nonumber
\end{eqnarray}
where $\Delta\omega=\omega_{0}-\omega_{_{\rm RF}}$ is the
frequency offset. Equation~(\ref{Hrot}) is obtained from
Eqs.~(\ref{H}) and~(\ref{HrotU}) for $ U =\exp(-i\omega_{_{\rm
RF}}  I_z t)$.

The initial state (before applying pulses) of the spin system at a
high temperature $T$ can be described by
\begin{eqnarray}
   \rho = Z^{-1} \exp(-\beta  {\cal H}_0) \approx  Z^{-1}(1-\beta  {\cal
  H}_0),
\label{rho0}
\end{eqnarray}
where $Z$ is the partition function and $\beta=1/(k_B T)$. The
term $\beta  {\cal H}_0/Z$ corresponds to a deviation density
matrix. Thus, the initial state $ \rho$ of a spin-$3/2$ system can
be approximated by
\begin{eqnarray}
   \rho \approx  \frac14(1 - \hbar\omega_0\beta  I_z),
\label{rho0b}
\end{eqnarray}
if $|\omega_Q|\ll |\omega_0|$.

The evolution of a state, given by $ \rho(t_0)$, during the
application of a single pulse of strength $\omega_1$ and duration
$t_p$ is described in the rotating frame by:
\begin{equation}
   \rho(t+t_p) =  {\cal U}(\omega_1,t_p)   \rho(t)
   {\cal U}^\dagger(\omega_1,t_p),
\label{U1}
\end{equation}
where the evolution operator is
\begin{equation}
   {\cal U}(\omega_1,t) = \exp[-(i/\hbar) {\cal H}_{\rm rot}
  t].
\label{U}
\end{equation}
The evolution of $ \rho(t)$ in the absence of pulses from the time
$t$ to $t+\Delta t$ is given by:
\begin{equation}
   \rho(t+\Delta t) =  {\cal U}(0,\Delta t)   \rho(t)  {\cal U}^\dagger(0,\Delta t). \label{U2}
\end{equation}
Analytical expressions for the evolution operator ${\cal
U}(\omega_1,t)$ and the corresponding density matrices can be
obtained by finding eigenvalues and eigenstates of ${\cal H} _{\rm
rot}$. For example, by assuming an RF pulse to be resonant with
the central line (i.e., $\omega_{_{\rm
RF}}=\omega_{12}=\omega_0$), and by setting $\phi=0$, we find the
following eigenvalues of ${\cal H} _{\rm rot}$:
\begin{equation}
  {\rm eig}({\cal H}_{\rm rot})=\left[
  \frac{\omega_1}2 + \Omega_{-},\frac{\omega_1}2 - \Omega_{-},
  -\frac{\omega_1}2 - \Omega_{+},-\frac{\omega_1}2 +
  \Omega_{+}\right]
\label{eigval}
\end{equation}
where $\Omega_{\pm}=\sqrt{\omega_1^2\pm\omega_1
\omega_Q+\omega^2_Q}$. The corresponding eigenvectors of ${\cal
H}_{\rm rot}$ for $m=1,2$ and $n=3,4$ are equal to:
\begin{eqnarray}
|V_{m}\rangle &=& {\cal N}_{m}[
  \sqrt{3}\omega_1(|3\rangle+|0\rangle)+
  y_{m}(|1\rangle+|2\rangle)],
  \nonumber\\
  |V_{n}\rangle &=& {\cal N}_n [
  \sqrt{3}\omega_1(|3\rangle-|0\rangle)+
  z_{n}(|1\rangle-|2\rangle)],
\label{eigvec}
\end{eqnarray}
where ${\cal N}_{m}$ and ${\cal N}_{n}$ are normalization
constants, and
\begin{eqnarray}
  y_m &=& \omega_1 + 2(-1)^m\Omega_{-}-2\omega_Q,
   \nonumber\\
  z_n &=& \omega_1 - 2(-1)^n \Omega_{+}+2\omega_Q.
  \label{eigvec2}
\end{eqnarray}
The general solution for ${\cal U}(\omega_1,t)$ is quite lengthy.
However by assuming that $|\omega_0|\gg |\omega_Q| \gg|\omega_1|$,
it can be effectively reduced to a form corresponding to all ideal
selective rotations as defined in Appendix~A. This can be shown by
expanding the elements of the matrix ${\cal U}(\omega_1,t)$ in a
power series of the parameter $\epsilon=|\omega_1|/|\omega_Q|$,
and, finally, keeping only the first term of this expansion.

For example, if the pulse is resonant with the central transition,
then the evolution operator ${\cal U}(\omega_1,t)$ can be
approximated by
\begin{equation}
  {\cal U}_{12}(\omega_1,t_p)= \MATT
 {\delta^*  & 0   & 0   & 0}
 {0  & \delta \cos(\omega_1 t_p)  & -i\delta \sin(\omega_1 t_p)  & 0}
 {0  & -i\delta \sin(\omega_1 t_p)  & \delta \cos(\omega_1 t_p)  & 0}
 {0  & 0   & 0   & \delta^*},
\label{Uapprox}
\end{equation}
where $\delta=\exp(i\omega_Q t_p)$. Note that ${\cal
U}_{12}(\omega_1,t)$ reduces to the perfect selective rotation
${\cal X}_{12}(\theta)= R^{(X)}_{12}(\theta)$, with
$\theta=2\omega_1 t_p$,  if the pulse duration is chosen such that
$\omega_Q t_p$ is a multiple of $2\pi$. Analogously, other
rotations $R^{(i)}_{mn}(\theta)$, given by Eq.~(\ref{A4}), can be
implemented for $i=X,Y,Z$ and $m,n=0,...,3$ with $m\neq n$.

\section{Implementing gates in spin-3/2 system}

Here, we discuss how to implement single- and two-qubit gates in
systems with spin-3/2. This can enable formally simple
implementations of arbitrary multi-qubit quantum algorithms by
applying sequences of NMR pulses in multi-level spin systems. We
focus on various NMR QST methods for a system with spin-3/2 nuclei
but our analysis can be easily generalized for larger spins.

Due to the Zeeman and quadrupolar interactions (shown in Fig.~1),
a spin-3/2 system is described in an external magnetic field by a
non-equidistant four-level energy spectrum. Thus, this system can
be referred to as a \emph{quartit} (also called ququart or
four-level qudit). The basic set of eigenfunctions of the system
can be described with the states $|mn\>\equiv|m\>_A |n\>_B$ of two
logical (or virtual) qubits $A$ and $B$ corresponding to an
ensemble of identical spin-1/2 pairs:
\begin{eqnarray}
|\fra32, \fra32\> &\equiv& |0\> \equiv |00\> , \quad
|\fra32,-\fra12\> \equiv |2\> \equiv |10\> ,
\nonumber \\
|\fra32, \fra12\> &\equiv& |1\> \equiv |01\> , \quad
|\fra32,-\fra32\> \equiv |3\> \equiv |11\> .\label{N01}
\end{eqnarray}
A pure state of a quartit can be written in this basis states as
\begin{eqnarray}
 \ket{\psi}&=& c_0 \ket{0}+ c_1 \ket{1}+c_2 \ket{2}+c_3 \ket{3}
 \label{N02}
\end{eqnarray}
in terms of the normalized complex amplitudes $c_i$, so an
arbitrary mixed state of a quartit is described by a density
matrix $\rho=[\rho_{nm}]_{4\times 4}$.

Our discussion in this section is based on a fundamental theorem
in quantum information according to which any quantum gate can be
constructed from single-qubit rotations and any nontrivial
two-qubit gate, e.g., the CNOT gate~\cite{SchleichBook}. In a
quartit, rotations of a virtual qubit $A$ and $B$, denoted,
respectively, by $R^A(\theta)$ and $R^B(\theta)$, can be
implemented by the application of two pulses:
\begin{eqnarray}
  R^A(\theta
  ) &=& R_{02}(\theta) R_{13}(\theta) ,
\nonumber \\
  R^B(\theta) &=& R_{01}(\theta) R_{23}(\theta) ,
\label{Q1}
\end{eqnarray}
where $R_{mn}(\theta)$ (with $R=X,Y,Z$) is a selective rotation
resonant with a transition between levels $|m\>$ and $|n\>$ as
defined in Appendix~A (see also Fig.~2).

Note that realizations of \emph{single} virtual qubit gates in a
qudit are more complicated than those for real qubits. In contrast
to those, usually \emph{two} virtual qubit gates can be realized
much simply, e.g., a CNOT-like gate can be implemented by applying
a \emph{single} $\pi$-pulse, e.g.,
\begin{eqnarray}
 S_{23}\equiv U'_{\rm CNOT} =\MATT
 {1 & 0 & 0 & 0}
 {0 & 1 & 0 & 0}
 {0 & 0 & 0 & -1}
 {0 & 0 & 1 & 0}
 = {\cal Y}_{23}(\pi).\quad
\label{Q2}
\end{eqnarray}
Similarly, a SWAP-like gate can also be implemented easily by
a \emph{single} $\pi$-pulse:
\begin{eqnarray}
  S_{12}\equiv U'_{\rm SWAP } &=& \MATT
 {1 & 0 & 0 & 0}
 {0 & 0 &-1 & 0}
 {0 & 1 & 0 & 0}
 {0 & 0 & 0 & 1}
 = {\cal Y}_{12}(\pi).
\label{Q3}
\end{eqnarray}
The above CNOT-like and SWAP-like gates can be related to the
standard CNOT and SWAP gates as follows:
\begin{eqnarray}
  U_{\rm CNOT}  &=& DU'_{\rm CNOT},
\nonumber \\
  U_{\rm SWAP} &=& U'_{\rm SWAP}D,
  \label{Q4}
\end{eqnarray}
where $D={\rm diag}([1,1,-1,1])$. One can define a SWAP-like gate
$S_{nm}$ between any levels $\ket{n}$ and $\ket{m}$ in a quartit,
simply as
\begin{equation}
  S_{nm} = {\cal Y}_{nm}(\pi),
 \label{Q5}
\end{equation}
which in special cases reduce to Eqs.~(\ref{Q2}) and~(\ref{Q3}).

It is worth noting that any unitary operator that can create
entanglement between a pair of qubits (or virtual qubits) is
universal. Thus, the standard SWAP gate $U_{\rm SWAP}$ is not
universal, as its entangling power is zero. In contrast to this
gate, the SWAP-like gate \emph{$U'_{\rm SWAP}$ is universal}, as
it can entangle qubits.

\section{Principles of $M_z$-based QST}

NMR quantum state tomography is a method for the complete
reconstruction of a given density matrix $\rho$ in a series of NMR
measurements. In general, to completely reconstruct a density
matrix $\rho$ for a quartit or two qubits, we need to determine 16
real parameters. Note that if the efficiency of a given detection
system is known then the 16th element can typically be found from
the normalization condition. Single NMR readout can only give some
elements of $\rho$: either diagonal (in case of $M_z$ detection)
or off-diagonal elements (for $M_{xy}$ detection), as discussed in
Appendix~B. The remaining matrix elements of the original density
matrix $\rho$ can be obtained by rotating it through properly
chosen rotational operations $R^{(k)}$, which change $\rho$ as
follows:
\begin{eqnarray}
  \rho^{(k)} &\equiv& R^{(k)} \rho (R^{(k)})^\dagger.
\label{N07}
\end{eqnarray}
These operations are performed before NMR readout measurements.
Thus, the reconstruction of a given density matrix is possible by
transforming $\rho$ through various rotations $R^{(k)}$ in such a
way that all the elements of $\rho$ go over into measurable ones
in a given detection method.

In the standard NMR $M_{xy}$ detection, one can directly determine
some of the off-diagonal elements of the density matrix. In
contrast to the $M_{xy}$ detection, one directly determines only
diagonal elements in the $M_z$ detection. It is worth noting that
the spectrum of a spin-3/2 system obtained via the $M_z$ detection
contains less information than the spectra obtained by the
$M_{xy}$ detection as discussed in Appendix~B: the $M_{xy}$
detection of a spin-3/2 system yields six real values, which
correspond to three peaks of real and those of imaginary parts of
the spectrum. Note that the $M_{xy}$ detection of a coupled two
spin-1/2 system can yield even more values if one could detect
signals from ensembles of two different spins simultaneously.

Tomography based on the measurements of the $M_{z}$ and
$M_{xy}$ magnetizations of spin-3/2 systems has been
performed in experiments reported in Refs.~\cite{Bonk04}
and~\cite{Kampermann05}, respectively.

\begin{figure}

 \fig{ \includegraphics[width=7cm]{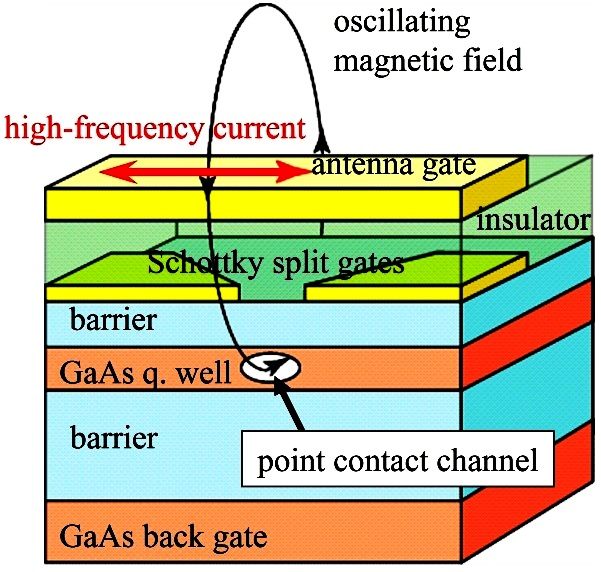}}

\caption{(Color online) Schematic diagram of the implementation of
a coherently and selectively controllable solid-state system with
nuclear spins-3/2 for quantum information processing and quantum
tomography. This  is an on-chip monolithic semiconductor device
integrated with a point-contact channel, fabricated by Yusa
\etal~\cite{Yusa05}. The point-contact channel (indicated with an
arrow), as defined by a pair of Schottky split gates, is composed
of a small ensemble of nuclear quadrupolar spins-3/2 of isotopes
$^{69}$Ga, $^{71}$Ga, and $^{75}$As. The nuclear spins in this
device can exhibit extremely-long decoherence times (a few
milliseconds). The antenna gate, which is insulated from the
Schottky gates, can locally irradiate the channel with an RF
field. The ensemble of spins-3/2 in the channel can be selectively
and coherently polarized and controlled by NMR techniques. The
resistance of the channel changes after an RF pulse, which
directly corresponds to the change of the longitudinal
magnetization $M_z$ induced by the change in the population of
nuclear-spin states. Thus, the discussed tomographic methods,
which are based on the detection of the longitudinal magnetization
of nuclear spins, can be effectively implemented in such devices.
This device is based on a quantum-well GaAs structure, but other
semiconductors can also be used. For example, by using the
semiconductor FeSb$_2$, instead of GaAs, the coherent control of
nuclear quadrupolar spins-7/2 for the isotope $^{123}$Sb would
enable quantum information processing with a quoctit (eight-level
qudit) corresponding to three virtual qubits. This figure is based
on Fig.~1(b) in Ref.~\cite{Yusa05}.}
 \label{fig3}
\end{figure}

\subsection{An implementation of $M_z$ detection in a
nanometer-scale device}

Here, we briefly describe an implementation of the NMR detection
of the longitudinal magnetization of a small ensemble of
quadrupolar spins-3/2, which is beyond the detection limits of
conventional NMR
techniques~\cite{Machida03,Yusa05,Hirayama06,Ota07,decoherence,Kondo08}.

This NMR detection was developed and applied in
Refs.~\cite{Yusa05} to an on-chip semiconductor device based on a
quantum-well structure shown in Fig.~3. This nanometer-scale
device is composed of a monolithic GaAs quantum well integrated
with a point contact channel and an antenna gate, where an
RF~field can be locally applied. The GaAs layer effectively forms
a two-dimensional electron gas. The point-contact channel is
composed of isotopes $^{69}$Ga, $^{71}$Ga, and $^{75}$As having
total ground-state spin $I$=3/2. The nuclear spins in the channel
can be selectively polarized by flowing current, while the spins
in the other regions are kept in thermal equilibrium. These
interactions between electron and nuclear spins are enhanced when
an external static magnetic field ${B}_0$ is applied to set the
system at the spin phase transition of the Landau level filling
factor 2/3~\cite{Hashimoto02}. The polarization is followed by
RF~pulses applied through the antenna gate, which enable
manipulation of the nuclear spins. This coherent manipulation
results in oscillations of the resistance of the point-contact
channel, which are directly related to the oscillations in the
longitudinal magnetization $M_z$. Reference~\cite{Yusa05} observed
clear oscillations reflecting all possible transitions between the
four nuclear-spin states (see Fig.~2) of each nuclide ($^{69}$Ga,
$^{71}$Ga, and $^{75}$As). This novel device, exhibiting extremely
low decoherence~\cite{Yusa05,decoherence}, opens new perspectives
to study characteristics of nuclear spins in nanoscale
semiconductors, but also to precisely control nuclear-spin states.
The arbitrary control of the superposition of the four spin-3/2
states enables the implementation of two-qubit coherent
operations~\cite{Hirayama06,Ota07}. Thus, the device offers new
possibilities to perform single- and two-qubit quantum gates, or
even to test simple quantum-information processing algorithms. A
fabrication of analogous device based on InAs and
InSb~\cite{Liu10}, instead of GaAs, where the isotopes $^{113}$In
and $^{115}$In have spin $I=9/2$ (a ten-level qudit) and the
isotope $^{123}$Sb has spin $I=7/2$ (an eight-level qudit), would
enable the implementation of three-qubit quantum gates and
algorithms. But it must be admitted that the devices are not
easily scalable for much higher number of virtual and/or real
qubits.

The initialization of the described device is relatively easy. We
can realize the effective pure state $\ket{3}$ by using
current-induced nuclear spin polarization with randomizing pulses
of $\omega_{01}$ and $\omega_{12}$. Once the state $\ket{3}$ is
realized, it is transferred to $\ket{0}$, $\ket{1},$ and $\ket{2}$
by applying a respective $\pi$ pulse as described by us in
Refs.~\cite{Hirayama06,Ota07}.

The estimated polarization of the nuclear spins is quite high.
Therefore, obtaining initial states with high purity should be
practically quite simple --- at least to start with the
pseudo-pure states $|0\>$ or $|4\>$.

There are many different possibilities to measure spectra. The
following is the simplest example applied in experiments described
in Ref.~\cite{Hirayama06}. After the preparation of a desired
state, we apply a pulse with a duration corresponding to a $\pi$
pulse and measure how a resistance changes as a function of
frequency. In case of the constant pulse-current amplitude, the
length of the $\pi$-pulse changes
$1/\sqrt{3}:1/\sqrt{4}:1/\sqrt{3}$ for $\omega_{01}$,
$\omega_{12}$ and $\omega_{23}$, respectively. (These differences
can be ignored in a simple experiment.) From this spectrum, we can
estimate a population difference between neighboring states, i.e.,
$\rho_{11}-\rho_{00}$, $\rho_{22}-\rho_{11}$, and
$\rho_{33}-\rho_{22}$ (see Sec.~VI.B). Another experimental
observation approach will be described in Sec.~VI.C.

\section{Linear reconstruction and the error robustness}

Various numerical procedures for reconstructing an unknown density
matrix $\rho$ from experimental data have been developed (see,
e.g., Refs.~\cite{ParisBook,DAriano03} and references therein).

The simplest and most intuitive QST is based on the inversion of a
linear system,
\begin{eqnarray}
  Ax = b,
\label{Axb}
\end{eqnarray}
where the real vector $x={\rm vec}(\rho)$ corresponds to the state
$\rho$ to be reconstructed. This vector can be defined in various
ways. Here, for a quartit state, we define $x$ as
\begin{equation}
x={\rm vec}(\rho) = [\rho_{00},{\rm Re} \rho_{01},{\rm Im}
\rho_{01},{\rm Re} \rho_{02},{\rm Im} \rho_{02}, ...,\rho_{33}]^T,
\label{Na1}
\end{equation}
where $\rho_{ij}$, for $i\le j$, are only included. Thus, a
density matrix $\rho$ can be expressed via the elements of the
vector $x$ as follows
\begin{eqnarray}
  \rho = \left[
\begin{array}{cccc}
 x_{1} & x_{2}+i x_{3} & x_{4}+i x_{5} & x_{6}+i x_{7} \\
 x_{2}-i x_{3} & x_{8} & x_{9}+i x_{10} & x_{11}+i x_{12} \\
 x_{4}-i x_{5} & x_{9}-i x_{10} & x_{13} & x_{14}+i x_{15} \\
 x_{6}-i x_{7} & x_{11}-i x_{12} & x_{14}-i x_{15} & x_{16} \\
\end{array}
\right]\!\!. \label{rho}
\end{eqnarray}
Moreover, in Eq.~(\ref{Axb}), ${b}$ is the \emph{observation
vector}, which contains the measured data; and $A$ is the
\emph{coefficient matrix}, which is also referred to as the
rotation matrix, or the data matrix in a more mathematical
context. Thus, the element $A_{j i}$ is the coefficient of $x_i$
in the $j$th equation ($j=1,...,N_{\rm eqs}$) for a chosen
measurement rotation. In our context, the observation vector
$b_{j}$ corresponds to the integrated area of the NMR spectra. The
number $N_{\rm eqs}$ of equations is given by $N_{\rm r}\times
N_{\rm vals}$, assuming $N_{\rm r}$ readouts (for a given
measurement), where each of them yields $N_{\rm vals}$ values
corresponding, e.g., to the number of peaks of an NMR spectrum
(including both real and imaginary parts). Usually, an extra
equation is added for the normalization condition, $\tr\rho=1$.
Thus, for a quartit, the observation vector has $N_{\rm eqs}$
elements and the coefficient matrix $A$ is of dimensions $N_{\rm
eqs}\times 16$.

Usually, there are more equations than unknowns. Such
\emph{overdetermined} problems can be solved as
\begin{eqnarray}
  {C} x=\tilde b,
\label{N09}
\end{eqnarray}
where ${C}\equiv[ {C}_{ij}]_{16 \times 16}={A}^\dagger {A}$ and
$\tilde b \equiv[\tilde b_{j}]_{16 \times 1}={A}^\dagger {b}$.
Equation~(\ref{N09}) results from the standard
least-squares-fitting analysis based on the minimalization of
$\chi^{2}=||Ax-b||^2$. Thus, one can easily calculate the solution
$x = {C}^{-1}\tilde b$ and, finally, reconstruct the sought
density matrix as
\begin{equation}
  \rho={\rm vec}^{-1}(x)={\rm vec}^{-1}\left({C}^{-1}\tilde b\right),
 \label{invX}
\end{equation}
as the inverse of Eq.~(\ref{Na1}).

Dozens of different linear-inversion-based QST protocols have been
proposed and applied (see, e.g.,~\cite{ParisBook,DAriano03} and
references citing those). Then the question arises: Which of them
are preferable for certain goals and tasks?

As an indicator of the quality of a linear-inversion-based QST
method, or more precisely its error robustness (or error
sensitivity), one can apply the so-called \emph{condition number}
defined as~\cite{AtkinsonBook,HighamBook,GolubBook}:
\begin{equation}
  {\rm cond}_{\alpha,\beta}(C) = \Vert C \Vert_{\alpha,\beta}\; \Vert C^{-1}
  \Vert_{\beta,\alpha}\ge 1,
\label{N11}
\end{equation}
where $C$ is a nonsingular square matrix and the convention is
used that ${\rm cond}_{\alpha,\beta}(C) = +\infty$ for a singular
matrix $C$. Moreover, $\Vert \cdot \Vert_{\alpha,\beta}$ denotes
the subordinate matrix norm, which can be defined via the vector
norms: $ \Vert C \Vert_{\alpha,\beta} = \max_{x\neq 0} \Vert Cx
\Vert_{\beta}/\Vert x \Vert_{\alpha}.$ Clearly, the condition
numbers depend on the applied norm. Here, we apply the spectral
norm only.

The spectral norm (also refereed to as the 2-norm) is given by the
largest singular value of $C$, i.e, $\Vert C \Vert_{2}\equiv
\Vert C \Vert_{2,2}=\max[{\rm svd}(C)]\equiv \sigma_{\max}(C)$,
where the function ${\rm svd}(C)$ gives the singular values of
$C$. Then this condition number is simply given by
\begin{eqnarray}
 \kappa(C)\equiv {\rm cond}_{22}(C)  = \frac{\sigma_{\max}(C)}{\sigma_{\min}(C)},
\label{kappa2}
\end{eqnarray}
where we have used $\Vert C^{-1} \Vert_{2}= \max[{\rm
svd}(C^{-1})]= \{\min[{\rm svd}(C)]\}^{-1}\equiv
\sigma^{-1}_{\min}(C).$ There are various geometrical, algebraic,
and physical interpretations of condition numbers (see
Ref.~\cite{Miran14} and references therein, in addition to
Refs.~\cite{AtkinsonBook,HighamBook,GolubBook,MeyerBook}). In
particular, according to the Gastinel-Kahan theorem, the inverse
of a condition number corresponds to the relative distance of a
nonsingular square matrix $C$ to the set of singular matrices.
Another, more physical interpretation can be given as
follows~\cite{AtkinsonBook}: Let us assume errors $\delta\tilde b$
in the observation vector $\tilde b$, which cause errors $\delta
x$ in the reconstructed vector $x$:
\begin{eqnarray}
  C(x+\delta x) &=& \tilde b+\delta \tilde b,
\label{Atkinson1}
\end{eqnarray}
then the following inequalities hold:
\begin{equation}
 \frac{1}{{\rm cond_{\alpha,\beta}}(C)}
 \frac{||\delta \tilde b||}{||\tilde b||} \le  \frac{||\delta x||}{||x||} \le {\rm cond_{\alpha,\beta}}(C)
 \frac{||\delta \tilde b||}{||\tilde b||}.
\label{Atkinson3}
\end{equation}
It is clear that when a condition number ${\rm
cond_{\alpha,\beta}}(C)\approx 1$, then small relative changes in
the observation vector $\tilde b$ cause equally small relative
changes in the reconstructed state $x$. This interpretation can be
generalized to include also errors $\delta C$ in the coefficient
matrix $C$.

Thus, by applying this general theorem, given in
Eq.~(\ref{Atkinson3}), to  QST, we can conclude that if ${\rm
cond}_{\alpha,\beta}(C)$ is small (large), then the coefficient
matrix $C$ and the corresponding QST method are called
\emph{well-conditioned} (\emph{ill-conditioned}), which means that
the method is robust (sensitive) to errors in the observation
vector $\tilde b$. For ill-conditioned QST, even a minor error in
$\tilde b$ can cause a large error in $x$. Some instructive
numerical examples of ill-conditioned problems are given in
Refs.~\cite{AtkinsonBook,Miran14}.

Condition numbers were applied to estimate the quality of optical
tomographic reconstructions in, e.g.,
Refs.~\cite{Bogdanov10,Miran14}. A condition number was also
calculated for the NMR tomography of two qubits (two
spins-1/2)~\cite{Roy10}. However, to our knowledge, these
parameters have not been applied yet to analyze the quality of QST
of any \emph{qudit} systems. More importantly, none of the
previous NMR tomographic methods exhibits the optimum robustness
against errors as described by a condition number equal or almost
equal to 1. Below we propose a few NMR QST protocols and compare
their error robustness based on the condition numbers to show that
some of our methods are optimal.

Note that the smallest singular value (or, equivalently,
eigenvalue) $\sigma_{\min}({C}) = \min[{\rm svd}({C})] =
||{C}^{-1}||_2$ of ${C}$ is also sometimes used as an
error-robustness parameter. This approach was applied in the
analysis of an NMR QST method in, e.g., Ref.~\cite{Long01}. In
comparison to $\sigma_{\min}({C})$, the condition numbers are much
better parameters of the error robustness as discussed in, e.g.,
Ref.~\cite{Miran14}.

\section{Observation approaches}

Here we specify three observation approaches based on the $M_z$
detection to be studied in detail in the next sections.

\subsection{Theoretical approach}

In an ideal $M_z$ detection, one can directly access all the
diagonal elements
\begin{eqnarray}
  b_n^{(k)} &=& \rho^{(k)}_{nn}
\label{OA1}
\end{eqnarray}
of any rotated density matrix $\rho^{(k)}\equiv R^{(k)}\rho\,
(R^{(k)})^{\dagger}$ for $k=1,...,N_{\rm r}$, where $N_{\rm r}$ is
the number of readouts (operations or sets of rotations). We refer
to this purely theoretical method as the \emph{theoretical
approach}.

\subsection{Ideal experimental approach}

In a more realistic observation approach, the information is
gathered from the $M_z$-spectra, where one can roughly estimate
the population differences ($\rho_{11}-\rho_{00}$,
$\rho_{22}-\rho_{11}$, and $\rho_{33}-\rho_{22}$) from the
amplitude of the signals by integrating the area of the peaks
centered at $\omega_{01}$, $\omega_{12}$, and $\omega_{23}$,
respectively. Thus, on including the normalization condition, we
have the following set of equations:
\begin{eqnarray}
  b_n^{(k)} &=& \rho^{(k)}_{n+1,n+1}-\rho^{(k)}_{nn}
  =  \tr(I_z^{(n+1,n)}\rho^{(k)}),
 \nonumber \\ 1 &=& \tr \rho^{(k)}
\label{OA2}
\end{eqnarray}
for each rotated density matrix $\rho^{(k)}$, where
$I_z^{(n+1,n)}=\ket{n+1}\bra{n+1}-\ket{n}\bra{n}$ is the
fictitious spin-1/2 operator for general spin. Note that
$b_n^{(k)}$ can be rescaled as $\bar b_n^{(k)}{\cal N}$, where the
constant ${\cal N}$ is usually chosen so the thermal equilibrium
magnetization vector is equal to a unit vector along the $z$
axis~\cite{LevittBook}. By referring to the ideal experimental
approach, we mean that based on Eq.~(\ref{OA2}).

Alternatively, the measured resistance in experiments performed
in, e.g., Refs.~\cite{Yusa05,Hirayama06,Ota07}, can be
proportional to the longitudinal magnetization $M_z \propto
\tr[\rho I_z]$ defined in terms of the total angular momentum
operator $I_z={\rm diag} ([\frac32,\frac12,-\frac12,-\frac32])$
for spin $I=3/2:$
\begin{equation}
  M_z^{(k)}\propto\tr[\rho^{(k)} I_z]=\frac{1}{2}
  (3\rho^{(k)}_{00}+\rho^{(k)}_{11}-\rho^{(k)}_{22}-3\rho^{(k)}_{33}).
\label{OA2extra}
\end{equation}
However, instead of studying this approach based on
Eq.~(\ref{OA2extra}), we apply a more practical observation method
based on the standard cyclically-ordered phase sequence (CYCLOPS)
technique.

\subsection{Non-ideal experimental approach using CYCLOPS}

Here, we study a practical measurement method by applying the
CYCLOPS to a $\pi/20$ reading pulse and
receiver~\cite{FreemanBook}. This method was used in, e.g., the
experiment on QST for quadrupolar nuclei of a liquid crystal by
Bonk \emph{et al.}~\cite{Bonk04}. In this observation approach,
the NMR spectra were obtained from free induction decay (FID)
averaged over each phase ($x,-y,-x,y$). This enables the
suppression of receiver imperfections and, thus, the cancellation
of artifacts from the NMR spectra. The intensities $b_{n}^{(k)}$
of the three ($n=1,2,3$) peaks of the averaged NMR spectrum for
the rotated deviation matrices
\begin{equation}
  \Delta\rho^{(k)}\equiv \rho^{(k)}-\tfrac14 {\cal I}, \quad \text{for}\;k=1,...,N_{\rm
r},
 \label{deviation}
\end{equation}
together with the normalization conditions are described by the
following set of equations for, e.g., the quartit:
\begin{eqnarray}
[b_{1}^{(k)},b_{2}^{(k)},b_{3}^{(k)}]^T&=&V {\rm
diag}(\Delta\rho^{(k)}), \nonumber \\ 0 &=& \tr
(\Delta\rho^{(k)}), \label{OA3}
\end{eqnarray}
where~\cite{Bonk04}:
\begin{equation}
V=\left[\begin{array}{cccc}
\sqrt{3}e_{11}e_{12} & - \sqrt{3}e_{12}e_{22} & - \sqrt{3}e_{23}e_{13} & - \sqrt{3}e_{13}e_{14}\\
2e_{13}e_{12}& 2e_{22}e_{23}& - 2e_{23}e_{22}& - 2e_{13}e_{12}\\
\sqrt{3}e_{13}e_{14} & \sqrt{3} e_{13}e_{23} & \sqrt{3}
e_{12}e_{22} & - \sqrt{3} e_{11}e_{12}
\end{array}\right].
\label{OA3a}
\end{equation}
The $n$th NMR peak corresponds to the transition between levels
$|n-1\rangle$ and $|n\rangle$. Above, ${\rm
diag}(\Delta\rho^{(k)})$ denotes a column vector of the diagonal
elements of $\Delta\rho^{(k)}$. The coefficients $e_{ij}$ are the
absolute values of the $\pi/20$ hard-reading pulse given by:
\begin{equation}
[e_{ij}]=\frac 14 \left[
\begin{array}{cccc}
 c_{31} & s_{zz} & c_{z,-z} & s_{3,-1} \\
 s_{zz} & c_{13} & s_{-1,3} & c_{z,-z} \\
 c_{z,-z} & s_{-1,3} & c_{13} & s_{zz} \\
 s_{3,-1} & c_{z,-z} & s_{zz} & c_{31}
\end{array}
\right], \label{eij}
\end{equation}
where $c_{xy}=x \cos \left(\frac{\pi }{40}\right)+y \cos
\left(\frac{3 \pi }{40}\right)$, $s_{xy}=x \sin \left(\frac{\pi
}{40}\right)+y \sin \left(\frac{3 \pi }{40}\right)$, and
$z=\sqrt{3}$. It seems that this method results in the coefficient
matrices, which are completely different from those obtained in
the ideal experimental approach. However, we will show that they
are practically very similar.
\begin{figure}
 \fig{ \includegraphics[width=7cm]{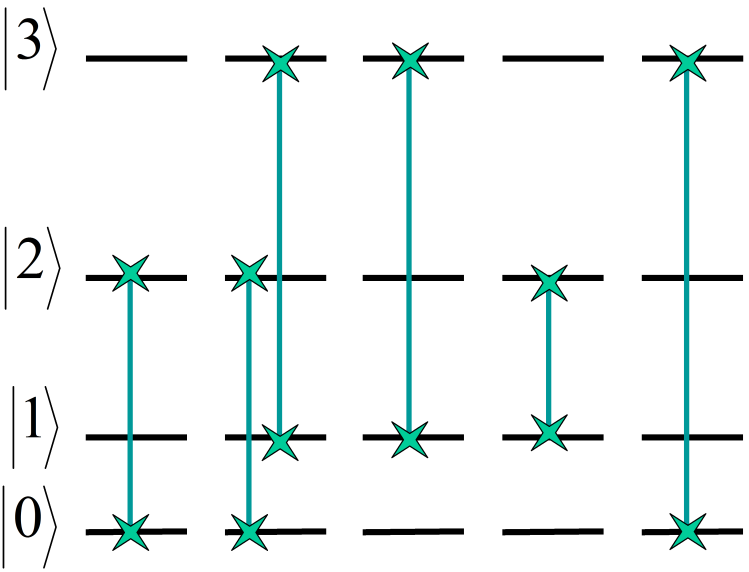}}
\caption{(Color online) Set of rotations $R^{\rm opt1}_{\rm
diag}$, given in Eq.~(\ref{Nfig4}), which enables the optimal
reconstruction of all the diagonal elements of $\rho$, by
measuring only the first peak of the $M_z$-spectra, i.e., the peak
corresponding to the transition $\ket{0}\leftrightarrow \ket{1}$.
The operation marked by $\times\!\!-\!\!\times$ denotes the
SWAP-like operation $S_{nm}\equiv {\cal Y}_{nm}(\pi)$ between the
levels $\ket{n}$ and $\ket{m}$. Reading pulses (which are not
shown here) are applied only after applying these SWAP-like gates.
The set $R^{\rm opt1}_{\rm diag}$ corresponds to the coefficient
matrix $A\equiv A^{\rm opt1}_{\rm diag}$, given in
Eq.~(\ref{Adiag_opt1}), yielding the smallest condition number
$\kappa(A^TA)=1$. Note that the last row in $A$ corresponds to the
normalization condition, while the first row corresponds to
applying a reading pulse without the SWAP. For brevity, this
trivial case corresponding to the identity operation is not
presented here.} \label{fig4}
\end{figure}
\begin{figure}
 \fig{ \includegraphics[width=7cm]{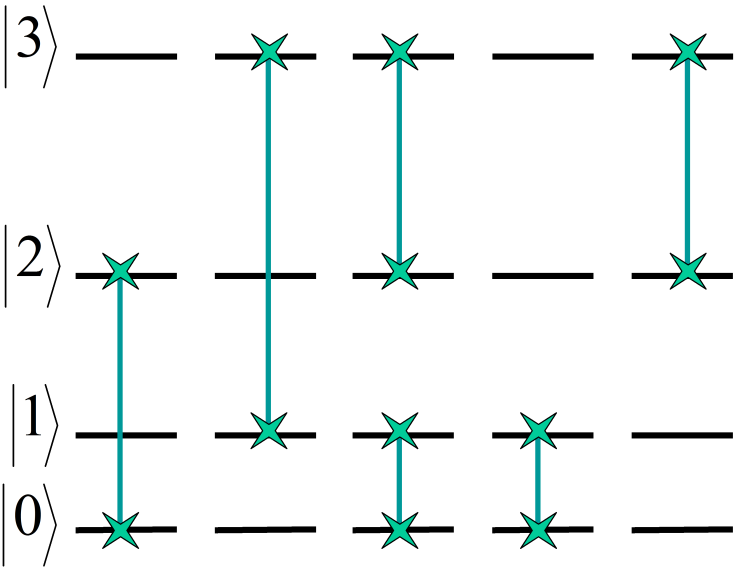}}
\caption{(Color online) Set of rotations $R^{\rm opt2}_{\rm
diag}$, given in Eq.~(\ref{Nfig5a}). Same as in Fig.~4, but the
coefficient matrix $A^{\rm opt2}_{\rm diag}$ corresponds to
measuring only the central peak of the $M_z$-spectra, i.e., that
corresponding to the transition $\ket{1}\leftrightarrow \ket{2}$.}
\label{fig5}
\end{figure}
\section{Optimal reconstruction of the diagonal elements of $\rho$}

Here we analyze the error robustness based on the condition number
$\kappa$ for the reconstruction of only diagonal terms $\rho_{nn}$
of a quartit density matrix $\rho$ for the three observation
approaches using various sets of rotations.

\subsection{Theoretical approach}

In the theoretical observation approach, we assume a direct access
to all the diagonal terms of $\rho$:
\begin{align}
  b^{(1)}_1 &= \rho_{00} \equiv x_1, \quad \;
  b^{(1)}_2 = \rho_{11} \equiv x_8,
\nonumber \\
  b^{(1)}_3 &= \rho_{22} \equiv x_{13}, \quad
  b^{(1)}_4 = \rho_{33} \equiv x_{16},
 \label{N35a}
\end{align}
where $\rho=\rho^{(1)}$. This implies that this partial tomography
is perfectly robust against errors, as described by the condition
number $\kappa=1$. Obviously, this robustness does not guarantee
that complete tomographic methods can also be perfectly robust
against errors.

\subsection{Ideal experimental  approach}

The set of equations~(\ref{OA2}) directly leads to the coefficient
matrices, which are in general different from those obtained by a
direct measurement of all the diagonal elements of $\rho$.
Nevertheless, from Eq.~(\ref{OA2}), one can easily determine all
the diagonal elements $\rho^{(k)}$, e.g., as follows:
\begin{eqnarray}
  \rho^{(k)}_{00} &=& \fra14- \fra14 (3 b^{(k)}_1 +2 b^{(k)}_2 +b^{(k)}_3),
\nonumber \\
  \rho^{(k)}_{11} &=& \fra14+ \fra14 (b^{(k)}_1 -2 b^{(k)}_2 -b^{(k)}_3),
\nonumber \\
  \rho^{(k)}_{22} &=& \fra14+ \fra14  (b^{(k)}_1 +2 b^{(k)}_2 -b^{(k)}_3),
\nonumber \\
  \rho^{(k)}_{33} &=& \fra14+ \fra14 (b^{(k)}_1 +2 b^{(k)}_2 +3 b^{(k)}_3).
\label{X8}
\end{eqnarray}
We can rewrite this problem in a matrix form, given by
Eq.~(\ref{Axb}), with its solution $x=(A^{\rm temp1}_{\rm
diag})^{-1}b$, where
\begin{eqnarray}
  A^{\rm temp1}_{\rm diag}&=&\left[
\begin{array}{llll}
  -1 & 1 & 0 & 0 \\
   0 &-1 & 1 & 0 \\
   0 & 0 &-1 & 1 \\
   1 & 1 & 1 & 1
\end{array}
   \right],
  \label{X9} \\
  x &=& [\rho_{00},\rho_{11},\rho_{22},\rho_{33}]^T,
  \label{X9a} \\
  b &=& [b_1^{(1)},b_2^{(1)},b_3^{(1)},1]^T,
  \label{X9b}
\end{eqnarray}
where, as usual, $b_n^{(1)}$ corresponds to the $n$th peak
resulting from the transition between the levels $\ket{n-1}$ and
$\ket{n}$ in the original matrix $\rho=\rho^{(1)}$. The condition
number reads $\kappa(A^TA)=6.83$, for $A\equiv A^{\rm temp1}_{\rm
diag}$. Thus, this direct application of the ideal experimental
observation method to reconstruct \emph{only} the diagonal terms
of $\rho$ for a quartit can magnify the relative error in the
observation vector $b$ by almost one order of magnitude.

Nevertheless, in the following, we show how to achieve
$\kappa(A^TA)=1$, even if the diagonal matrix elements are not
directly measured. It is worth noting that the condition number
$\kappa(A^TA)$ can also be equal to 1 for analogous $M_z$-based
QST of the diagonal elements of a density matrix for two
spatially-separated qubits. This is because the diagonal matrix
elements can be directly measured, so no reconstruction of these
elements is required.

Note that this coefficient matrix $A^{\rm temp1}_{\rm diag}$ is
unbalanced, which implies that some elements of $\rho$ are
measured more often than others. Specifically, there are only two
nonzero elements in the first and last columns of $A^{\rm
temp1}_{\rm diag}$ (corresponding to $\rho_{00}$ and $\rho_{33}$)
and three nonzero elements in the other columns of $A$
(corresponding to $\rho_{11}$ and $\rho_{22}$). To overcome this
problem, let us apply the pulse $S_{13}\equiv {\cal Y}_{13}(\pi)$,
which corresponds to the SWAP-like gate. Then, we measure only the
first peak (corresponding to the transition
$\ket{0}\leftrightarrow \ket{1}$) of the rotated density matrix
$S_{13}\rho S_{13}^\dagger$. Thus, by adding this equation to
$A^{\rm temp1}_{\rm diag}$, one obtains
\begin{equation}
  A^{\rm temp2}_{\rm diag}=[A^{\rm temp1}_{\rm diag};(-1,0,0,1)].
 \label{X10}
\end{equation}
Then the condition number becomes $\kappa(A^TA)=2$ for $A=A^{\rm
temp2}_{\rm diag}$, which is much smaller than that for $A=A^{\rm
temp1}_{\rm diag}$. One can then obtain a more balanced
coefficient matrix $A^{\rm opt}_{\rm diag}$ by adding two
equations to $A^{\rm temp2}_{\rm diag}$, which correspond to the
first peak of the rotated density matrices $S_{12}\rho
S_{12}^\dagger$ and $S_{03}\rho S_{03}^\dagger$. Thus, we have
\begin{eqnarray}
  A_{\rm diag}^{\rm opt1} &=& \begin{pmatrix}
    -1 & 1 & 0 & 0 \\
    0  &1 & -1 & 0 \\
    0  & 0 &-1 & 1 \\
    -1 & 0 & 0 & 1 \\
    -1 & 0 & 1 & 0 \\
    0  & 1 & 0 &-1 \\
    s  & s & s & s \
  \end{pmatrix}.
\label{Adiag_opt1}
\end{eqnarray}
Note that we have multiplied the second row in Eq.~(\ref{X9}) by
the factor (-1) to obtain Eq.~(\ref{Adiag_opt1}), which enables us
to slightly simplify the following Eq.~(\ref{Nfig4}). This
operation does not affect the corresponding condition numbers. As
usual, the last row (equation) in Eq.~(\ref{Adiag_opt1})
corresponds to the normalization condition, where $s$ is the
scaling factor, which is set here as $s=1$. Then we find that
$C_{\rm diag}^{\rm opt}=(A_{\rm diag}^{\rm opt1})^\dagger A_{\rm
diag}^{\rm opt1}=4I_4,$ where $I_4$ is the four-dimensional
identity operator. In general, this factor $s$ determines the
contribution of the last equation to the whole set $Ax=b$ of
equations and can be chosen such that $\kappa$ is minimized.

Thus, we have shown that the condition number $\kappa(C_{\rm
diag}^{\rm opt})=1$ indicates the optimality of the coefficient
matrix $A_{\rm diag}^{\rm opt}$ for the ideal experimental
observation approach. This matrix $A_{\rm diag}^{\rm opt}$ can be
obtained from the following set of rotations:
\begin{equation}
  R^{\rm opt1}_{\rm diag}=[I,S_{02},S_{13}S_{02},S_{13},S_{12},S_{03}],
 \label{Nfig4}
\end{equation}
as shown in Fig.~4, using the SWAP-like gates $S_{nm}\equiv {\cal
Y}_{nm}(\pi)$. Here we assume that only the first peak is
measured, while the other two peaks are ignored, in the
$M_z$-spectra of the rotated density matrices
$\rho^{(k)}=R^{(k)}\rho\,(R^{(k)})^\dagger$ for all the rotations
$R^{(k)}$ in Eq.~(\ref{Nfig4}).

Alternatively, we can swap the elements $\rho_{nn}$ in such a way
that only the central peak, corresponding to
$\ket{1}\leftrightarrow \ket{2}$, is measured. Namely, one can use
the following set of rotations
\begin{eqnarray}
  R^{\rm opt2}_{\rm diag}&=&[I,S_{02},S_{13},S_{01}S_{23},S_{01},S_{23}]
  \label{Nfig5a}
\end{eqnarray}
as shown in Fig.~5, which leads to the following coefficient
matrix
\begin{eqnarray}
  A_{\rm diag}^{\rm opt2} &=& \left(
\begin{array}{cccc}
 0 & -1 & 1 & 0 \\
 1 & -1 & 0 & 0 \\
 0 & 0 & 1 & -1 \\
 -1 & 0 & 0 & 1 \\
 -1 & 0 & 1 & 0 \\
 0 & -1 & 0 & 1 \\
 s & s & s & s \\
\end{array}
\right), \label{Adiag_opt2}
\end{eqnarray}
where the last row corresponds, as usual, to the normalization
condition. By setting $s=1$, one finds that $C_{\rm diag}^{\rm
opt}=(A_{\rm diag}^{\rm opt2})^\dagger A_{\rm diag}^{\rm
opt2}=4I_4,$ the same as for $(A_{\rm diag}^{\rm opt1})^\dagger
A_{\rm diag}^{\rm opt1}$. This property holds since $A_{\rm
diag}^{\rm opt1}$ and $A_{\rm diag}^{\rm opt2}$ differ only in the
order of their rows and in the opposite sign of all the elements
of some of their rows. Therefore, these coefficient matrices can
be considered equivalent,
\begin{equation}
  A_{\rm diag}^{\rm opt1}\cong A_{\rm diag}^{\rm opt2}.
 \label{equivalent1}
\end{equation}
Moreover, the two-photon rotations, given in Eq.~(\ref{Nfig5a}),
can be replaced by single-photon transitions, e.g., $S_{02}$ can
be replaced by $S_{01}S_{12}$, and $S_{13}$ by
$S_{12}S_{23}S_{12}$, as will be described in general terms in
Sec.~X. This might be an advantage from the experimental point of
view.

It is worth noting that the state vector $x$ is defined as in
Eq.~(\ref{X9a}) for both methods,  based on the rotations $R^{\rm
opt1}_{\rm diag}$ and $R^{\rm opt2}_{\rm diag}$. However, the
observation vectors $b$ are defined as follows: (i) For the
rotations $R^{\rm opt1}_{\rm diag}$, one measures
\begin{eqnarray}
  b &=& [b_1^{(1)},b_1^{(2)},...,b_1^{(6)},s]^T,
\label{b1}
\end{eqnarray}
where $b_1^{(k)}$ corresponds to the first peak obtained for the
rotated density matrix $\rho^{(k)}=R^{\rm opt1}_{{\rm
diag},k}\rho\,(R^{\rm opt1}_{{\rm diag},k})^\dagger$. (ii) For the
rotations $R^{\rm opt2}_{\rm diag}$, the observation vector reads
\begin{eqnarray}
  b &=& [b_2^{(1)},b_2^{(2)},...,b_2^{(6)},s]^T,
\label{b2}
\end{eqnarray}
where $b_2^{(k)}$ corresponds to the second peak obtained for
$\rho^{(k)}=R^{\rm opt2}_{{\rm diag},k}\rho\,(R^{\rm opt2}_{{\rm
diag},k})^\dagger$.

\subsection{Non-ideal experimental approach using CYCLOPS}

The set of Eqs.~(\ref{OA3}) can be rewritten in the matrix form
$Ax=b$, where
\begin{eqnarray}
  A^{\rm temp3}_{\rm diag} &=& [V;(s,s,s,s)],
\label{X15} \\
  x &=& [\rho_{00}-\tfrac{1}{4},\rho_{11}-\tfrac14,\rho_{22}-\tfrac14,\rho_{33}-\tfrac14]^T,\quad\quad
\label{X15b}\\
  b &=& [b_1^{(1)},b_2^{(1)},b_3^{(1)},0]^T.
\label{X15c}
\end{eqnarray}
Then, we find the condition number to be $\kappa(A^{T}A)=98.46$,
if $A=A^{\rm temp3}_{\rm diag}$ and $s=1$. It means that the
determination of all the diagonal terms of $\rho$ for a quartit
using the standard CYCLOPS method can be relatively sensitive to
errors. Indeed, the relative errors in the observation vector $b$
can be magnified in the reconstructed vector $x$ by almost two
orders of magnitude. Thus, one could conclude that the
reconstruction of all (not only diagonal) elements of $\rho$ can
be worse by at least two orders of magnitude, in comparison to the
corresponding QST methods but assuming the ideal experimental
observation approach. In contrast to these tentative conclusions,
we will show below that, in fact, the error robustness can be
described by the condition number $\kappa\approx 1$ for both
partial and complete tomographic methods.

Analogously to the theoretical approach, we can also optimize the
set of rotations of $\rho$ and measure only some peaks of the
$M_z$-spectra if some non-ideal experimental observation approach
is applied. Here we analyze the optimization of rotations for the
CYCLOPS method.

First, we optimize the value of the scaling factor $s$ in $A\equiv
A^{\rm temp3}_{\rm diag}$. By choosing $s\in(0.1,0.25)$, we find
that $\kappa(A^{T}A)=6.1375$, which is almost one order smaller
than $\kappa(A^{T}A)$ for $s=1$. Now, we apply the sets of
rotations $R^{{\rm opt}1}_{\rm diag}$ and $R^{{\rm opt}2}_{\rm
diag}$ to obtain the coefficient matrices $\bar A^{{\rm
opt}1}_{\rm diag}$ and $\bar A^{{\rm opt}2}_{\rm diag}$ assuming
the scaling factors $s=0.2318$ and $s=0.3043$, respectively.
Specifically, the optimal value of $s$ for a given coefficient
matrix $A$ is chosen here as $\max_{i,j}A_{ij}'$, where $A'$ is
the matrix $A$ but without the last row (i.e., with the nonzero
elements equal to $s$). Note that this last equation is added to
include the normalization condition for measuring the $n$th peak
($n=1,2$) using the CYCLOPS method. Although our precise
expressions for the coefficient matrices $\bar A^{{\rm opt}1}_{\rm
diag}$ and $\bar A^{{\rm opt}2}_{\rm diag}$ are quite lengthy, and
thus not shown here, we find that
\begin{equation}
  \bar A_{\rm diag}^{\rm opt1} \approx A_{\rm diag}^{\rm opt1}\cong A_{\rm diag}^{\rm opt2} \approx \bar A_{\rm diag}^{\rm opt2},
 \label{equivalent2}
\end{equation}
where Eq.~(\ref{equivalent1}) was used. Our precise calculations
result in the following condition numbers
\begin{eqnarray}
  \kappa(A^TA) &=& 1.0371 \quad {\rm for}\; A=\bar A^{\rm opt1}_{\rm diag},
\nonumber \\
  \kappa(A^TA) &=& 1.0384 \quad {\rm for}\; A=\bar A^{\rm opt2}_{\rm
  diag},
\label{kkk1}
\end{eqnarray}
which are very close to one.

\section{Reconstruction of the off-diagonal elements of
$\rho$}

Now we propose several sets of rotations for the reconstruction of
all the off-diagonal terms $\rho_{nm}$ (with $n\neq m$) for the
three observation approaches and study the robustness of these
methods against errors.

\subsection{Theoretical approach}

Our first temporary proposal for QST of a spin-3/2 system is based
on a natural choice of 12 rotations (see also Fig.~6):
\begin{eqnarray}
  R^{\rm temp}_{\rm offdiag} = [Y_{01},X_{01},Y_{12},X_{12},Y_{23},X_{23},\nonumber \\
  Y_{02},X_{02},Y_{13},X_{13},Y_{03},X_{03}],
\label{Nfig6}
\end{eqnarray}
where hereafter $X_{mn}\equiv {\cal X}_{mn}(\halfpi)$ and
$Y_{mn}\equiv {\cal Y}_{mn}(\halfpi)$ as special cases of the
selective rotations ${\cal X}_{mn}(\theta)$ and ${\cal
Y}_{mn}(\theta)$ defined in Appendix~A. Thus, the method is based
on 6 single-photon, 4 two-photon, and 2 three-photon transitions.

In the $M_z$ detection approach, we can determine the diagonal
elements $[\rho_{00}, \rho_{11} , \rho_{22}, \rho_{33}]$ of a
density matrix $\rho$. By denoting the diagonal elements of
$\rho^{(k)}$ as ${\rm diag}(\rho^{(k)}) \equiv
(\rho_{nn}^{(k)})_n$, the following elements:
\begin{eqnarray}
{\rm
diag}(\rho^{(1)})&=&[f_{01}^{(22)},f_{01}^{(00)},\rho_{22},\rho_{33}],
\nonumber \\
{\rm
diag}(\rho^{(2)})&=&[f_{01}^{(13)},f_{01}^{(31)},\rho_{22},\rho_{33}],
\nonumber \\
{\rm
diag}(\rho^{(3)})&=&[\rho_{00},f_{12}^{(22)},f_{12}^{(00)},\rho_{33}],
\nonumber \\
{\rm
diag}(\rho^{(4)})&=&[\rho_{00},f_{12}^{(13)},f_{12}^{(31)},\rho_{33}],
\nonumber \\
{\rm
diag}(\rho^{(5)})&=&[\rho_{00},\rho_{11},f_{23}^{(22)},f_{23}^{(00)}],
\nonumber \\
{\rm
diag}(\rho^{(6)})&=&[\rho_{00},\rho_{11},f_{23}^{(13)},f_{23}^{(31)}],
\label{Nfig6rho} \\
{\rm
diag}(\rho^{(7)})&=&[f_{02}^{(22)},\rho_{11},f_{02}^{(00)},\rho_{33}],
\nonumber \\
{\rm
diag}(\rho^{(8)})&=&[f_{02}^{(13)},\rho_{11},f_{02}^{(31)},\rho_{33}],
\nonumber \\
{\rm
diag}(\rho^{(9)})&=&[\rho_{00},f_{13}^{(22)},\rho_{22},f_{13}^{(00)}],
\nonumber \\
{\rm
diag}(\rho^{(10)})&=&[\rho_{00},f_{13}^{(13)},\rho_{22},f_{13}^{(31)}],
\nonumber \\
{\rm diag}(\rho^{(11)})
&=&[f_{03}^{(22)},\rho_{11},\rho_{22},f_{03}^{(00)}],
\nonumber \\
{\rm diag}(\rho^{(12)})
&=&[f_{03}^{(13)},\rho_{11},\rho_{22},f_{03}^{(31)}], \nonumber
\end{eqnarray}
are found for the set of rotations given by Eq.~(\ref{Nfig6}),
where the auxiliary function $f_{mn}^{(kl)}$ is defined by
\begin{eqnarray}
  f_{mn}^{(kl)} &=& \frac 12 (\rho_{mm}+i^k \rho_{mn}+i^l \rho_{nm}+
  \rho_{nn}).
\label{N22}
\end{eqnarray}
In the theoretical approach, all these equations can determine the
coefficient matrices ${A}=A^{\rm temp}_{\rm offdiag}$, which are
based on the set of $N_{\rm eqs}=48$ equations given by
Eq.~(\ref{OA1}). The singular values of ${C}=A^TA$ are found to be
${\rm svd}(C)=\{12,8^{\otimes 3},2^{\otimes 12}\}$, where our
compact notation $\sigma_i^{\otimes n}$ denotes that $\sigma_i$
occurs $n$ times. Thus, we can determine the condition number
describing the error robustness of the QST method as
$\kappa(C)=6$, which is clearly not optimal. However, by analyzing
Eq.~(\ref{Nfig6rho}), one can find that all the off-diagonal
elements of $\rho$ can directly be determined as follows:
\begin{align}
   &2x_2=f_{01}^{(00)}-f_{01}^{(22)},
   &2x_3=f_{01}^{(31)}-f_{01}^{(13)},
\nonumber \\ \nonumber
   &2x_9=f_{12}^{(00)}-f_{12}^{(22)},
   &2x_{10}=f_{12}^{(31)}-f_{12}^{(13)},
\\ \nonumber
&2x_{14}=f_{23}^{(00)}-f_{23}^{(22)},
&2x_{15}=f_{23}^{(31)}-f_{23}^{(13)},
\label{x_offdiag}\\
 &2x_{4}=f_{02}^{(00)}-f_{02}^{(22)},
 &2x_{5}=f_{02}^{(31)}-f_{02}^{(13)},
\\ \nonumber
 &2x_{11}=f_{13}^{(00)}-f_{13}^{(22)},
 &2x_{12}=f_{13}^{(31)}-f_{13}^{(13)},
\\ \nonumber
 &2x_{6}=f_{03}^{(00)}-f_{03}^{(22)},
 &2x_{7}=f_{03}^{(31)}-f_{03}^{(13)}.
\end{align}
Thus, the corresponding condition number can be decreased to 1.

\subsection{Ideal experimental approach}

The $M_z$-based tomography for the rotations, given by
Eq.~(\ref{Nfig6}), in the ideal experimental observation approach
can be understood as follows: When we apply the $Y_{01}$ pulse
after a certain photon operation, we obtain the diagonal
components including $\rho_{01}$ and $\rho_{10}$. In our $M_z$
detection method, one of the three signals, which corresponds to
the differences of the populations between four spin states, is
proportional to $\RE(\rho_{01})+\RE(\rho_{10})$. Because
$\rho_{mn}=\rho_{nm}^*$, we can obtain
$\RE(\rho_{01})=\RE(\rho_{10})$. Similarly, $Y_{12}$, $Y_{23}$,
$Y_{02}$, $Y_{13}$, and $Y_{03}$ give us other elements
$\RE(\rho_{mn})=\RE(\rho_{nm})$. Imaginary parts are also
estimated by applying the $X_{mn}$ pulse by noting that
$\IM(\rho_{mn})=-\IM(\rho_{nm})$. Although QST needs a few
multiphoton operations, this method looks simple and easy to
interpret.

In the ideal experimental approach, corresponding to
Eq.~(\ref{OA2}), we obtain ${C}=A^TA$, where $A=A^{\rm temp}_{\rm
offdiag}$, having the following singular values:
\begin{eqnarray}
{\rm svd}(C)=\{48., 24.25, 16.17, 9.97, 6., 5.45, 5^{\otimes
2},\nonumber \\ 4.91,  4.37, 3^{\otimes 2}, 2.92, 2.26, 2, 1.71\},
\end{eqnarray}
which yield the condition number $\kappa(C)=28.14$, which is far
from being optimal.

However, by analyzing the equations in Eq.~(\ref{Nfig6rho}), one
can conclude that (at least) some of the off-diagonal terms of
$\rho$ can be measured directly, i.e.,
\begin{eqnarray}
  &b_1^{(1)}= 2x_2,   \quad\;    b_1^{(2)}=2 x_3,\, \quad\;\; b_2^{(3)}= 2x_9,
  \nonumber \\
  &b_2^{(4)}=2 x_{10},\quad b_3^{(5)}=2 x_{14}, \quad\; b_3^{(6)}= 2x_{15},
 \label{M7}
\end{eqnarray}
where $b_n^{(k)}$ corresponds to the $n$th peak of the $M_z$
spectrum obtained in the CYCLOPS method for the rotated density
matrix $\rho^{(k)}=R_k\rho R_k^\dagger$, where the rotation $R_k$
is given by the $k$th element in Eq.~(\ref{Nfig6}).

In order to directly measure other off-diagonal elements of
$\rho$, one can swap some quartit levels, say $\ket{k}$ and
$\ket{l}$, by applying the $\pi$-pulse $S_{kl}\equiv{\cal
Y}_{kl}(\pi)$. For example, one can use the following set of
rotations:
\begin{eqnarray}
R^{\rm opt0}_{\rm offdiag}  &=&
[Y_{01},X_{01},Y_{12},X_{12},Y_{23},X_{23},Y_{01}S_{12},X_{01}S_{12}, \nonumber\\
&&Y_{12}S_{23},X_{12}S_{23},Y_{01}S_{13},X_{01}S_{13}],\quad\quad
 \label{Nfig7}
\end{eqnarray}
as shown in Fig.~7.  Then, all the off-diagonal terms can be
directly measured including
\begin{eqnarray}
  &b_1^{(7)}=- 2x_4,\quad\quad b_1^{(8)}=-2 x_5, \quad  b_2^{(9)}= -2x_{11}, \nonumber \\
  & b_2^{(10)}=-2x_{12},\quad b_1^{(11)}= -2x_{6}, \quad b_1^{(12)}=-2x_{7},
 \label{M9}
\end{eqnarray}
in addition to those given in Eq.~(\ref{M7}). Thus, the
corresponding coefficient matrix becomes
\begin{eqnarray}
A^{\rm opt0}_{\rm offdiag}  =
2\left(
\begin{array}{cccccccccccc}
 1 & 0 & 0 & 0 & 0 & 0 & 0 & 0 & 0 & 0 & 0 & 0 \\
 0 & 1 & 0 & 0 & 0 & 0 & 0 & 0 & 0 & 0 & 0 & 0 \\
 0 & 0 & 0 & 0 & 0 & 0 & 1 & 0 & 0 & 0 & 0 & 0 \\
 0 & 0 & 0 & 0 & 0 & 0 & 0 & 1 & 0 & 0 & 0 & 0 \\
 0 & 0 & 0 & 0 & 0 & 0 & 0 & 0 & 0 & 0 & 1 & 0 \\
 0 & 0 & 0 & 0 & 0 & 0 & 0 & 0 & 0 & 0 & 0 & 1 \\
 0 & 0 & \bar1 & 0 & 0 & 0 & 0 & 0 & 0 & 0 & 0 & 0 \\
 0 & 0 & 0 & \bar1 & 0 & 0 & 0 & 0 & 0 & 0 & 0 & 0 \\
 0 & 0 & 0 & 0 & 0 & 0 & 0 & 0 & \bar1 & 0 & 0 & 0 \\
 0 & 0 & 0 & 0 & 0 & 0 & 0 & 0 & 0 & \bar1 & 0 & 0 \\
 0 & 0 & 0 & 0 & \bar1 & 0 & 0 & 0 & 0 & 0 & 0 & 0 \\
 0 & 0 & 0 & 0 & 0 & \bar1 & 0 & 0 & 0 & 0 & 0 & 0 \\
\end{array}
\right), \label{Aoffdiag0}
\end{eqnarray}
where $\bar1=-1$. The reconstructed vector for all the
off-diagonal elements reads
\begin{equation}
x=[x_{2}, x_{3}, x_{4}, x_{5}, x_{6}, x_{7}, x_{9}, x_{10},
x_{11}, x_{12}, x_{14}, x_{15}]^T,
\end{equation}
while the observation vector is
\begin{eqnarray}
b&=&[
  b_1^{(1)},
  b_1^{(2)},
  b_2^{(3)},
  b_2^{(4)},
  b_3^{(5)},
  b_3^{(6)},
\nonumber \\ &&
  b_1^{(7)},
  b_1^{(8)},
  b_2^{(9)},
  b_2^{(10)},
  b_1^{(11)},
  b_1^{(12)}]^T,
\end{eqnarray}
as implied by Eqs.~(\ref{M7}) and~(\ref{M9}). This observation
vector is obtained by measuring only a properly-chosen single peak
in a given $M_z$ spectrum, while the other two peaks are ignored.
Specifically, to determine a chosen term $b_n^{(k)}$, from those
in Eqs.~(\ref{M7}) and~(\ref{M9}), one should only measure the
$n$th peak of the $M_z$-spectra corresponding to the transition
$\ket{n-1}\leftrightarrow \ket{n}$ of the rotated density matrix
$\rho^{(k)}=R_k\rho R_k^\dagger$, where $R_k=R^{\rm opt0}_{{\rm
offdiag},k}$.

We can operationally simplify the problem by requiring that always
the same $n$th peak is measured in the all $M_z$-spectra. For
example, to measure always the first peak, one can perform the
SWAP-like operations. Thus, we propose the following optimal (in
terms of $\kappa=1$) set of rotations
\begin{eqnarray}
R^{\rm opt1}_{\rm offdiag} &=&
[Y_{01},X_{01},S_{02}Y_{12},S_{02}X_{12},Y_{01}S_{13}S_{02},\nonumber \\
&&X_{01}S_{13}S_{02},Y_{01}S_{12},X_{01}S_{12},S_{02}Y_{12}S_{23},\nonumber\\
&&S_{02}X_{12}S_{23},Y_{01}S_{13},X_{01}S_{13}],
 \label{Nfig8}
\end{eqnarray}
as shown in Fig.~8, which corresponds to the following coefficient
matrix
\begin{eqnarray}
A^{\rm opt1}_{\rm offdiag}  = 2\left(
\begin{array}{cccccccccccc}
 1 & 0 & 0 & 0 & 0 & 0 & 0 & 0 & 0 & 0 & 0 & 0 \\
 0 & 1 & 0 & 0 & 0 & 0 & 0 & 0 & 0 & 0 & 0 & 0 \\
 0 & 0 & 0 & 0 & 0 & 0 & \bar1 & 0 & 0 & 0 & 0 & 0 \\
 0 & 0 & 0 & 0 & 0 & 0 & 0 & \bar1 & 0 & 0 & 0 & 0 \\
 0 & 0 & 0 & 0 & 0 & 0 & 0 & 0 & 0 & 0 & 1 & 0 \\
 0 & 0 & 0 & 0 & 0 & 0 & 0 & 0 & 0 & 0 & 0 & 1 \\
 0 & 0 & \bar1 & 0 & 0 & 0 & 0 & 0 & 0 & 0 & 0 & 0 \\
 0 & 0 & 0 & \bar1 & 0 & 0 & 0 & 0 & 0 & 0 & 0 & 0 \\
 0 & 0 & 0 & 0 & 0 & 0 & 0 & 0 & 1 & 0 & 0 & 0 \\
 0 & 0 & 0 & 0 & 0 & 0 & 0 & 0 & 0 & 1 & 0 & 0 \\
 0 & 0 & 0 & 0 & \bar1 & 0 & 0 & 0 & 0 & 0 & 0 & 0 \\
 0 & 0 & 0 & 0 & 0 & \bar1 & 0 & 0 & 0 & 0 & 0 & 0 \\
\end{array}
\right). \label{Aoffdiag1}
\end{eqnarray}
In this approach only the first peak is measured. Alternatively,
we can swap the quartit levels in such a way that only the central
peak is measured. Then, the optimal tomography can be achieved for
the following set of rotations
\begin{eqnarray}
R^{\rm opt2}_{\rm offdiag} &=&
[Y_{12}S_{02},X_{12}S_{02},Y_{12},X_{12},Y_{12}S_{13},\nonumber \\
&&X_{12}S_{13},Y_{12}S_{01},X_{12}S_{01},Y_{12}S_{23},\nonumber \\
&&X_{12}S_{23},Y_{12}S_{01}S_{23},X_{12}S_{01}S_{23}],
 \label{Nfig9}
\end{eqnarray}
as shown in Fig.~9, which corresponds to the following coefficient
matrix
\begin{eqnarray}
A^{\rm opt2}_{\rm offdiag}  = 2\left(
\begin{array}{cccccccccccc}
 1 & 0 & 0 & 0 & 0 & 0 & 0 & 0 & 0 & 0 & 0 & 0 \\
 0 & \bar1 & 0 & 0 & 0 & 0 & 0 & 0 & 0 & 0 & 0 & 0 \\
 0 & 0 & 0 & 0 & 0 & 0 & 1 & 0 & 0 & 0 & 0 & 0 \\
 0 & 0 & 0 & 0 & 0 & 0 & 0 & 1 & 0 & 0 & 0 & 0 \\
 0 & 0 & 0 & 0 & 0 & 0 & 0 & 0 & 0 & 0 & \bar1 & 0 \\
 0 & 0 & 0 & 0 & 0 & 0 & 0 & 0 & 0 & 0 & 0 & 1 \\
 0 & 0 & 1 & 0 & 0 & 0 & 0 & 0 & 0 & 0 & 0 & 0 \\
 0 & 0 & 0 & 1 & 0 & 0 & 0 & 0 & 0 & 0 & 0 & 0 \\
 0 & 0 & 0 & 0 & 0 & 0 & 0 & 0 & \bar1 & 0 & 0 & 0 \\
 0 & 0 & 0 & 0 & 0 & 0 & 0 & 0 & 0 & \bar1 & 0 & 0 \\
 0 & 0 & 0 & 0 & \bar1 & 0 & 0 & 0 & 0 & 0 & 0 & 0 \\
 0 & 0 & 0 & 0 & 0 & \bar1 & 0 & 0 & 0 & 0 & 0 & 0 \\
\end{array}
\right). \label{Aoffdiag2}
\end{eqnarray}
The advantage of the rotations $R^{\rm opt2}_{\rm offdiag}$, in
comparison to $R^{\rm opt1}_{\rm offdiag}$, resides in the lower
number of two-photon rotations. In addition, the remaining
two-photon SWAP-like rotations $S_{02}$ and $S_{13}$, listed in
Eq.~(\ref{Nfig9}), can be replaced by a sequence of single-photon
rotations as described in Sec.~X.

It is seen that
\begin{equation}
  A^{\rm opt0}_{\rm offdiag} \cong A^{\rm opt1}_{\rm offdiag} \cong A^{\rm opt2}_{\rm
  offdiag},
 \label{equivalent3}
\end{equation}
are equivalent up to an irrelevant multiplication of some of their
rows by the factor (-1).

In conclusion, we find that the proposed sets of rotations for the
reconstruction of all the off-diagonal density-matrix elements in
this ideal observation approach are optimal, as leading to the
lowest value of the condition number
\begin{equation}
  \kappa \left[(A^{{\rm opt},l}_{\rm offdiag})^T A^{{\rm opt},l}_{\rm offdiag}\right]
  = 1,
 \label{kappa_offdiag}
\end{equation}
for $l=0,1,2$.

\begin{figure}

 \fig{ \includegraphics[width=7cm]{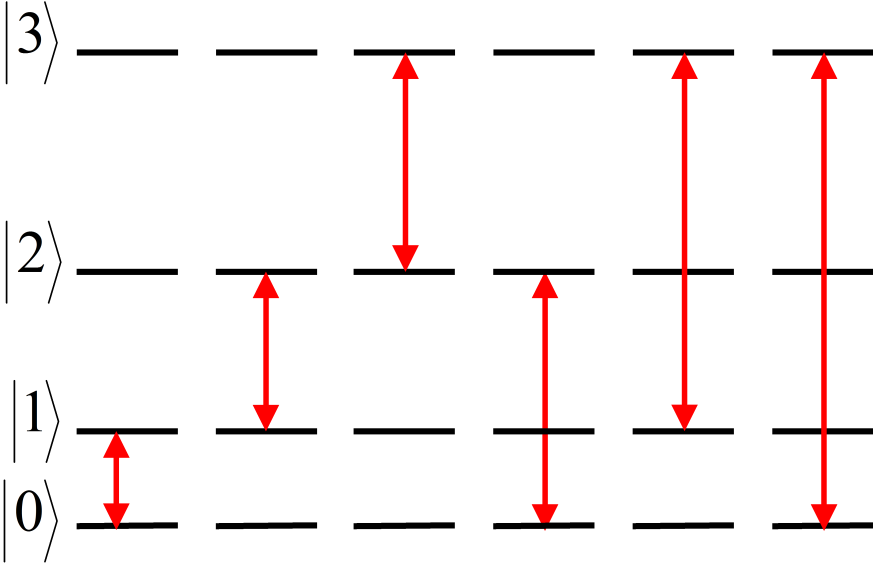}}

\caption{(Color online) Set of rotations $R^{\rm temp}_{\rm
offdiag}$, given in Eq.~(\ref{Nfig6}), for the reconstruction of
all the off-diagonal elements of $\rho$ in the 12 series of
measurements by applying the $\pi/2$-pulses: $X_{nm}={\cal
X}_{nm}(\pi/2)$ and $Y_{nm}={\cal Y}_{nm}(\pi/2)$, which are in
both cases marked by the red double arrows. In the theoretical
approach, the off-diagonal elements can directly be obtained by
Eq.~(\ref{x_offdiag}), which implies that the method is optimal
with $\kappa=1$. However, this is not the optimal method assuming
the experimental approaches, where only some of the off-diagonal
elements can directly be measured, according Eq.~(\ref{M7}), while
other elements are calculated indirectly, which results in
$\kappa>1$.} \label{fig6}
\end{figure}

\begin{figure}
 \fig{ \includegraphics[width=7cm]{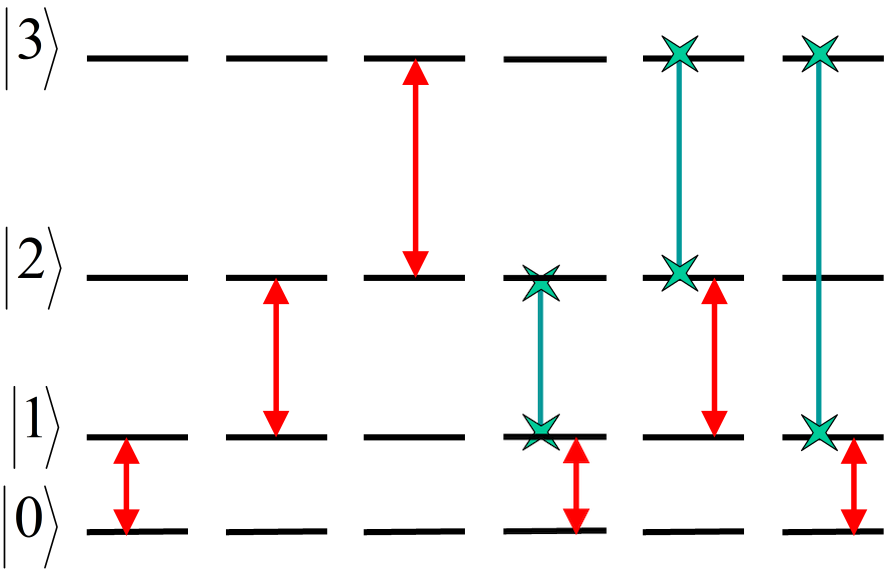}}
\caption{(Color online) Set of rotations $R^{\rm opt0}_{\rm
offdiag}$, given in Eq.~(\ref{Nfig7}) for the optimal
reconstruction of all the off-diagonal elements of $\rho$. The
optimal coefficient matrix $A^{\rm opt0}_{\rm offdiag}$ is
obtained by measuring only a properly-chosen single peak of each
$M_z$ spectrum, as indicated in Eqs.~(\ref{M7}) and~(\ref{M9}). }
\label{fig7}
\end{figure}

\begin{figure}
 \fig{ \includegraphics[width=7cm]{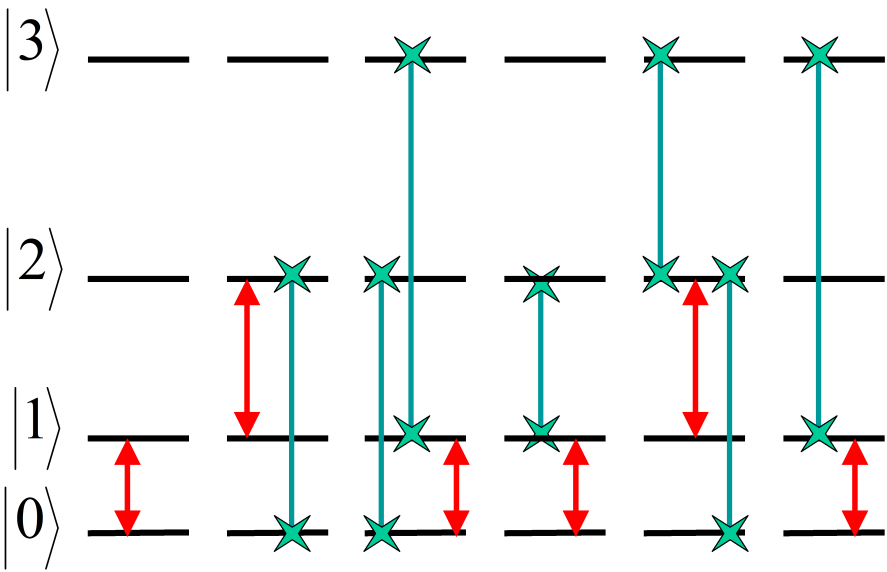}}
\caption{(Color online) Set of rotations $R^{\rm opt1}_{\rm
offdiag}$, given in Eq.~(\ref{Nfig8}) for the optimal
reconstruction of all the off-diagonal elements of $\rho$. Same as
in Fig.~6, but the optimal matrix $A^{\rm opt1}_{\rm offdiag}$ is
obtained by measuring only the first peak of the $M_z$-spectra.}
\label{fig8}
\end{figure}

\begin{figure}
 \fig{ \includegraphics[width=7cm]{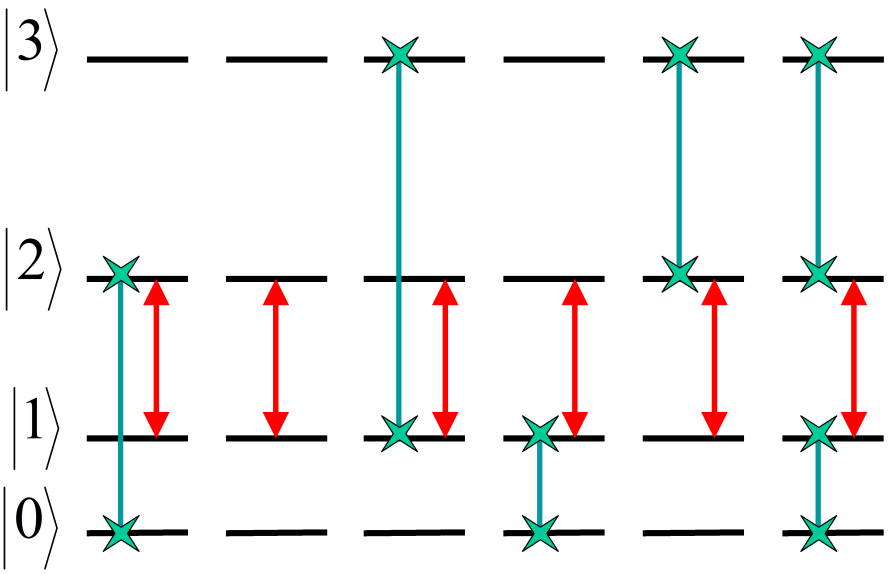}}
\caption{(Color online) Set of rotations $R^{\rm opt2}_{\rm
offdiag}$, given in Eq.~(\ref{Nfig9}). Same as in Fig.~8, but the
optimal matrix $A^{\rm opt2}_{\rm offdiag}$ is obtained by
measuring only the central peak of the $M_z$-spectra.}
\label{fig9}
\end{figure}
\subsection{Non-ideal experimental approach via CYCLOPS}

The experimental observation approach based on the CYCLOPS method
can be considered an imperfect (or noisy) version of the above
ideal experimental approach based on Eq.~(\ref{OA2}). For example,
the first peak $b_1^{(1)}$ in the $M_z$ spectrum after the
rotation $Y_{01}$ corresponds to:
\begin{equation}
  b_1^{(1)} = c_1x_1-c_2x_2+c_8x_8-c_{13}x_{13}-c_{16}x_{13}
  \approx -c_2 x_2,
 \label{1row}
\end{equation}
where $c_{1}\approx c_{8}\approx 0.0014$, $c_{2}\approx 0.4608$,
$c_{13}\approx 0.0029$, and $c_{16}\approx 9\times 10^{-16}$. Note
that the element $x_2$ is dominant, at least, by three orders of
magnitude in comparison to the other elements, as $c_2\approx
161\times \max_{i\neq 2}c_i$. Then, we find that the condition
number $\kappa(A^{\rm opt}_{\rm offdiag})\approx 1.000$ for the
rotations $R^{{\rm opt},l}_{\rm offdiag}$ (for $l=0,1,2$) assuming
the CYCLOPS measurement. This approximate calculation is performed
by ignoring the contributions of the diagonal terms $x_1$, $x_8$,
$x_{13}$, and $x_{16}$.

\section{Optimal reconstruction of all the elements of $\rho$}

\subsection{Theoretical approach}

By analyzing Eqs.~(\ref{N35a}) and~(\ref{x_offdiag}), all of the
16 elements of $x$ (and, thus, $\rho$) can be accessed directly in
this theoretical observation approach. Therefore, the
corresponding coefficient matrix is diagonal, and the
reconstruction of $x$ is trivial. Thus, this QST is perfectly
robust against errors, as the condition number is equal to 1.

\subsection{Ideal experimental approach}

In order to reconstruct all the elements of $\rho$ we can combine
the optimal reconstructions for the off-diagonal elements (based
on the optimal coefficient matrix $A_{\rm offdiag}^{{\rm opt},k}$
corresponding to the rotations $R_{\rm offdiag}^{{\rm opt},k}$ for
$k=0,1,2$) and diagonal elements (described by $A_{\rm diag}^{{\rm
opt},l}$ corresponding to $R_{\rm diag}^{{\rm opt},l}$  for
$l=1,2)$. For example, the combined coefficient matrices of the
dimensions $19\times 16$ can read
\begin{eqnarray}
 A_{\rm opt1}&=&[A_{\rm offdiag}^{\rm opt1};A_{\rm diag}^{\rm
opt1}],
\nonumber \\
  A_{\rm opt2}&=&[A_{\rm offdiag}^{\rm opt2};A_{\rm diag}^{\rm
opt2}], \label{Afull2}
\end{eqnarray}
corresponding to the sets of rotations
\begin{eqnarray}
  R_{\rm opt1}&=&[R_{\rm offdiag}^{\rm opt1};R_{\rm diag}^{\rm
opt1}],
\nonumber \\
  R_{\rm opt2}&=&[R_{\rm offdiag}^{\rm opt2};R_{\rm diag}^{\rm
opt2}], \label{Rfull2}
\end{eqnarray}
respectively. The last row in the matrices in Eq.~(\ref{Afull2})
corresponds to the normalization condition with the scaling factor
$s=1$. Thus, we obtain the total optimal coefficient matrix
\begin{eqnarray}
A_{\rm opt1}=
\begin{pmatrix}
 0 \, 2 \, 0 \, 0 \, 0 \, 0 \, 0 \, 0 \, 0 \, 0 \, 0 \, 0 \, 0 \, 0 \, 0 \, 0 \\
 0 \, 0 \, 2 \, 0 \, 0 \, 0 \, 0 \, 0 \, 0 \, 0 \, 0 \, 0 \, 0 \, 0 \, 0 \, 0 \\
 0 \, 0 \, 0 \, \bar2 \, 0 \, 0 \, 0 \, 0 \, 0 \, 0 \, 0 \, 0 \, 0 \, 0 \, 0 \, 0 \\
 0 \, 0 \, 0 \, 0 \, \bar2 \, 0 \, 0 \, 0 \, 0 \, 0 \, 0 \, 0 \, 0 \, 0 \, 0 \, 0 \\
 0 \, 0 \, 0 \, 0 \, 0 \, 0 \, 0 \, 0 \, 0 \, 0 \, 0 \, 0 \, 0 \, 2 \, 0 \, 0 \\
 0 \, 0 \, 0 \, 0 \, 0 \, 0 \, 0 \, 0 \, 0 \, 0 \, 0 \, 0 \, 0 \, 0 \, 2 \, 0 \\
 0 \, 0 \, 0 \, \bar2 \, 0 \, 0 \, 0 \, 0 \, 0 \, 0 \, 0 \, 0 \, 0 \, 0 \, 0 \, 0 \\
 0 \, 0 \, 0 \, 0 \, \bar2 \, 0 \, 0 \, 0 \, 0 \, 0 \, 0 \, 0 \, 0 \, 0 \, 0 \, 0 \\
 0 \, 0 \, 0 \, 0 \, 0 \, 0 \, 0 \, 0 \, 0 \, 0 \, 2 \, 0 \, 0 \, 0 \, 0 \, 0 \\
 0 \, 0 \, 0 \, 0 \, 0 \, 0 \, 0 \, 0 \, 0 \, 0 \, 0 \, 2 \, 0 \, 0 \, 0 \, 0 \\
 0 \, 0 \, 0 \, 0 \, 0 \, \bar2 \, 0 \, 0 \, 0 \, 0 \, 0 \, 0 \, 0 \, 0 \, 0 \, 0 \\
 0 \, 0 \, 0 \, 0 \, 0 \, 0 \, \bar2 \, 0 \, 0 \, 0 \, 0 \, 0 \, 0 \, 0 \, 0 \, 0 \\
 \bar1 \, 0 \, 0 \, 0 \, 0 \, 0 \, 0 \, 1 \, 0 \, 0 \, 0 \, 0 \, 0 \, 0 \, 0 \, 0 \\
 0 \, 0 \, 0 \, 0 \, 0 \, 0 \, 0 \, 1 \, 0 \, 0 \, 0 \, 0 \, \bar1 \, 0 \, 0 \, 0 \\
 0 \, 0 \, 0 \, 0 \, 0 \, 0 \, 0 \, 0 \, 0 \, 0 \, 0 \, 0 \, \bar1 \, 0 \, 0 \, 1 \\
 \bar1 \, 0 \, 0 \, 0 \, 0 \, 0 \, 0 \, 0 \, 0 \, 0 \, 0 \, 0 \, 0 \, 0 \, 0 \, 1 \\
 \bar1 \, 0 \, 0 \, 0 \, 0 \, 0 \, 0 \, 0 \, 0 \, 0 \, 0 \, 0 \, 1 \, 0 \, 0 \, 0 \\
 0 \, 0 \, 0 \, 0 \, 0 \, 0 \, 0 \, 1 \, 0 \, 0 \, 0 \, 0 \, 0 \, 0 \, 0 \, \bar1 \\
 s \, 0 \, 0 \, 0 \, 0 \, 0 \, 0 \, s \, 0 \, 0 \, 0 \, 0 \, s \, 0 \, 0 \, s \\
\end{pmatrix}
, \label{Acomplete}
\end{eqnarray}
where $\bar{1}=-1$ and $\bar{2}=-2$. For brevity, the analogous
coefficient matrix $A_{\rm opt2}$ is not presented here. A simple
calculation shows that $C_{{\rm opt},l}=(A_{{\rm opt},l})^\dagger
A_{{\rm opt},l}$ (for $l=1,2$) is proportional to the identity
operator, so the condition number $\kappa(C_{{\rm opt},l})=1$.
Thus, the proposed tomographic methods are optimal concerning
their robustness against errors assuming the ideal experimental
observations.

\subsection{Non-ideal experimental approach using CYCLOPS}

In the non-ideal experimental approach, we can follow the analysis
for the ideal experimental approach. In particular, we can apply
Eq.~(\ref{Acomplete}), but for the properly chosen scaling
factors. It should be stressed that Eq.~(\ref{Acomplete}) is only
an approximation of $\bar A_{\rm opt1}$ for the non-ideal case.
For example, the first row for exact $\bar A_{\rm opt1}$ is given
by Eq.~(\ref{1row}). However, we observe that
\begin{equation}
   \bar A_{\rm opt1} \approx A_{\rm opt1} \cong A_{\rm opt2}  \approx \bar A_{\rm opt2},
 \label{AAAfull}
\end{equation}
where the combined coefficient matrices $\bar A_{{\rm opt},l}$
(for $l=1,2)$ are obtained for the sets of rotations $R_{{\rm
opt},l}$, given in Eq.~(\ref{Rfull2}), analogously to $A_{{\rm
opt},l}$, given in Eq.~(\ref{Afull2}). Note that the last row in
$\bar A_{{\rm opt}1}$ and $\bar A_{{\rm opt}2}$ corresponds to the
normalization condition with the scaling factors $s=0.2304$ and
$s=0.3043$, respectively. Moreover, in the solution $x=A^{-1}b$,
the observation vector $b$ is equal to
$[b_l^{(1)},b_l^{(2)},...,b_l^{(18)},0]^T$ for $l=1,2$,
respectively, and the reconstructed state vector
$x=[x_1,x_2,...,x_{16}]^T-\frac14$ is related to $\rho$ by
Eq.~(\ref{Na1}). By performing precise numerical calculations, we
conclude that the condition numbers are very close 1, i.e.,
\begin{eqnarray}
  \kappa(A^TA) &=& 1.0592\;\approx\; 1 \quad {\rm for}\; A=\bar A_{\rm opt1},
\nonumber \\
  \kappa(A^TA) &=& 1.0528\;\approx\; 1 \quad {\rm for}\; A=\bar A_{\rm opt2}.
\label{kkk2}
\end{eqnarray}
Thus, even the non-ideal QST, as based on the CYCLOPS method, can
be almost perfectly robust to errors, as described by their
condition numbers $\kappa$.

\section{Single-photon replacements for multiphoton rotations}

The error-robustness analysis is based on the properties of the
coefficient matrices $A$ and, thus, enables to find experimental
setups for the reliable QST even without specific experimental
data. The optimization in our approach resides in replacing
degenerate multiphoton (multi-quantum) rotations by single-photon
ones.

For example, some of the discussed sets of rotations for QST
include single-photon $X$~rotations ($X_{01}, X_{12}, X_{23}$),
two-photon $X$~rotations ($X_{02}, X_{13}$), and a three-photon
$X$~rotation ($X_{03}$) together with analogous $Y$~rotations.
Especially the three-photon transitions are not the simplest to be
realized experimentally due to the degeneracy between
$\omega_{03}/3$ and $\omega_{12}$ if the second order quadrupolar
shifts are neglected (see Fig.~1). Namely, we want to perform the
three-photon rotations $Y_{03}$ and $X_{03}$ between levels
$\ket{0}$ and $\ket{3}$ (for brevity, we say the 0-3 rotation)
solely without changing populations between levels $\ket{1}$ and
$\ket{2}$. We can effectively rotate 1-2 without rotating 0-3, but
we are not able to rotate 0-3 without rotating 1-2. So a feasible
tomographic method should be described without direct rotations
0-3. Under this requirement, it is easy to show analytically that
one needs combinations of at least two rotations for some of the
operations for complete reconstruction. Then, unfortunately, the
above interpretation of the tomographic operations, given by
Eq.~(\ref{Nfig6rho}), loses its clarity.

First, we calculate replacements for multiphoton $X$~rotations. By
inspection, we find that
\begin{eqnarray}
{\cal X}_{0n}(\theta ) &=&S_{1n}{\cal X}_{01}(\theta ){S}^\dag
_{1n}  \nonumber \\
&=&{S}_{01}\,{\cal X}_{1n}(-\theta )\,{S}^\dag_{01}  \nonumber \\
&=&{S}_{02}\,{\cal X}_{2n}(-\theta )\,{S}^\dag_{02}  \label{N23} \\
&=&{S}_{2n}\,{\cal X}_{02}(\theta )\,{S}^\dag_{2n}  \nonumber \\
&=&{S}_{01}\,{S}_{2n}\,{\cal X}_{12}(-\theta
)\, {S}^\dag _{2n}\,{S}^\dag_{01}, \nonumber
\end{eqnarray}
given in terms of SWAP-like operations $S_{kl}\equiv{\cal
Y}_{kl}(\pi).$ Note that ${\cal Y}_{k,n}^{T}(\theta ) ={\cal
Y}_{k,n}^{\dag }(\theta )={\cal Y} _{k,n}(-\theta )$, so
$S^{\dag}_{kl}={\cal Y}_{kl}(-\pi).$ Analogously, other
replacements can be found. By repeatedly applying the first
relation in Eq.~(\ref{N23}) we obtain a general formula for any
two levels $k<n-1$:
\begin{eqnarray}
{\cal X}_{kn}(\theta )=S_{k+1,n}\,{\cal X}_{k,k+1}(\theta)\,{S}^\dag
_{k+1,n}. \label{N24}
\end{eqnarray}
Alternatively, one can apply the second relation in Eq.~(\ref{N23})
to obtain:
\begin{eqnarray}
{\cal X}_{kn}(\theta ) ={S}^{(k,n)}\,{\cal X}_{n-1,n}[(-1)^{n-k+1}\theta
]\,(S^{(k,n)})^\dag, \label{N25}
\end{eqnarray}
where
\begin{eqnarray}
{S}^{(k,n)}\equiv {S}_{k,k+1}\,{S}_{k+1,k+2} \cdots{S}_{n-2,n-1}.
\label{N26}
\end{eqnarray}
In the same way, we find replacements for multiphoton
$Y$~rotations for any $\theta$ and $k<n-1$:
\begin{eqnarray}
{\cal Y}_{kn}(\theta ) &=& S_{k+1,n}\,{\cal Y}_{k,k+1}(\theta
)\,S^\dag_{k+1,n}
\nonumber \\
&=& {S}^{(k,n)}\,{\cal Y}_{n-1,n}[(-1)^{n-k+1}\theta]\,(S^{(k,n)})^\dag,
\label{N27}
\end{eqnarray}
in terms of the pulse sequences given by Eq.~(\ref{N26}).

In a special case, for a given QST method of the quartit system,
the $X$~rotations based on three-photon (0-3) and two-photon (0-2
and 1-3) transitions can be replaced by various sequences of
rotations requiring only single-photon transitions, e.g.,
\begin{eqnarray}
{\cal X}_{03}(\theta) &=& {S}_{01}\, {S }_{23}\, \,{\cal
X}_{12}(-\theta) \,{S}_{23}^\dag\, {S}_{01}^\dag,
\nonumber \\
{\cal X}_{02}(\theta) &=&{S}_{01} \,{\cal X}_{12}(-\theta)\,
{S}_{01}^\dag, \label{N28}
 \\
{\cal X}_{13}(\theta) &=& {S}_{12}\, {\cal X}_{23}(-\theta) \,
{S}_{12}^\dag, \nonumber
\end{eqnarray}
and analogously for the multi-quantum $Y$~rotations.

Finally, we point out some practical aspects in the described
realization of a nanometer-scale device in a relation to the
problem of degeneracy between $\omega_{03}/3$ and $\omega_{12}$.
The rotation frequency is proportional to the first Bessel
function of the oscillation field strength for the coherent
rotation between levels $\ket{1}$ and $\ket{2}$, but proportional
to the third Bessel function for that between $\ket{0}$ and
$\ket{3}$. Therefore, the 0-3 rotation becomes negligible if the
applied field is weak. Moreover, it is possible to select the
oscillating field strength, which satisfies some angle rotation
for 0-3, which differs from a multiple of 2$\pi$ rotation for 2-3.
Therefore, it is possible to realize a pure 0-3 operation without
2-3 rotation. However, current amplitude necessary for this
operation might be very high and the operation is not realistic
from the view point of heating. Another QST method based on
sequences of the two-photon pulses $X_{02}$, $X_{13}$, $Y_{02}$
and $Y_{13}$ would also be experimentally feasible. But usually
rotations between the closest levels are much faster and easier to
perform.

Here, we give a simple solution to omit rotations requiring
three-photon transitions in, e.g., the rotations $R^{\rm
temp}_{\rm offdiag}$ is to express them as combinations of three
one-photon and two-photon rotations as described above. However,
we find that combinations of only two rotations are usually
sufficient. Thus, we suggest the following three-photon rotations
$Y_{03}\rightarrow Y_{01} S_{13}$ and $X_{03}\rightarrow X_{01}
S_{13}$. Note that the new operations do not require the rotation
0-3. We mention that the two-photon transitions can also be
replaced by single-photon transitions with the help of the
sequences of rotations given by Eq.~(\ref{N28}).

\section{Conclusions}

We described various methods for implementing quantum state
tomography for systems of quadrupolar nuclei with spin-3/2
(equivalent to quartit) in an unconventional approach to NMR,
which is based on the measurement of longitudinal magnetization
$M_z$ instead of the standard measurement of the transverse
magnetization $M_{xy}$~\cite{SlichterBook}.

This work has been motivated by the demonstration of
high-precision $M_z$-based NMR techniques of coherent manipulation
of nuclear spins $I=3/2$ ($^{69}$Ga, $^{71}$Ga, and $^{75}$As) in
a GaAs quantum-well device based on an the fractional quantum Hall
effect~\cite{Yusa05}. The device, exhibiting extremely low
decoherence~\cite{decoherence}, offers new possibilities to study
interactions in semiconductors but also enables the realization of
single- and two-qubit quantum gates~\cite{Hirayama06} and,
possibly, testing simple quantum-information processing
algorithms.

Although our presentation of the protocols of QST of large-nuclear
systems was focused on the nanoscale semiconductor device of
Ref.~\cite{Yusa05}, it should be stressed that these protocols can
also be readily applied to large-nuclear spins in other systems.

We proposed methods with optimized sets of rotations. The
optimization was applied in order to improve the robustness
against errors, as quantified by condition numbers.

Some of the proposed QST methods for a quartit system require the
three-photon transitions (between levels $|0\rangle$ and
$|3\rangle$), which are induced by relatively strong pulses.
Unfortunately, such pulses can simultaneously induce transitions
between levels $|1\rangle$ and $|2\rangle$. Thus, from a practical
point of view, it is desired to apply only weak pulses selectively
inducing single-photon transitions. We showed how the rotations
requiring multiphoton transitions can be replaced by combinations
of rotations based only on single-photon transitions.

By applying the condition number based on the spectral
norm~\cite{HighamBook}, we compared robustness against errors in
the measured data for all the described tomographic methods. We
have assumed three observation approaches corresponding to: (i) an
ideal $M_z$ detection, where all the diagonal elements $\rho_{nn}$
($n=0,...,3$) of a density matrix can be directly accessed; (ii)
an ideal experimental $M_z$ detection, where the population
differences ($\rho_{11}-\rho_{00}$, $\rho_{22}-\rho_{11}$, and
$\rho_{33}-\rho_{22}$) can be estimated from the amplitude of the
signals by integrating the area of the peaks centered at
$\omega_{01}$, $\omega_{12}$, and $\omega_{23}$ (see Fig.~2),
respectively; and (iii) the non-ideal (``noisy'') experimental
detection based on the CYCLOPS method, where the information
gathered from the $M_z$-spectra corresponds to some linear
functions of the diagonal elements $\rho_{nn}$, as given by
Eq.~(\ref{OA3}).

For the QST methods for a quartit (i.e., two virtual qubits) using
the experimental approaches (including the CYCLOPS method), the
condition number $\kappa$ is either exactly equal to 1 or very
close to 1. This means that the proposed methods are optimally
robust against errors.

Let us now compare the error robustness of the discussed NMR QST
methods with some known optical QST methods (see, e.g.,
Ref.~\cite{Miran14} for a review) for two physically-distinct
qubits: The well-known QST protocol of James \etal~\cite{James01},
which is solely based on local measurements, yields the condition
number $\kappa=60.1$. The QST of Refs.~\cite{Altepeter05,Burgh08}
is based on the standard separable basis composed of all of the 36
two-qubit eigenstates of the tensor products of the Pauli
operators. This often-applied QST yields $\kappa=9$. Another QST,
which is based on local measurements of the 16 tensor products of
the Pauli operators, yields $\kappa=2$. In contrast to these
optical methods, only the recently-proposed QST of
Ref.~\cite{Miran14}, which was also experimentally
demonstrated~\cite{Bartkiewicz15}, is optimal since it yields the
condition number $\kappa=1$. This tomography of two optical qubits
is based on local and global measurements of generalized Pauli
operators. It is worth noting that our optimal NMR tomography is
based on a smaller set of measurements in comparison to that for
the optimal optical tomography~\cite{Miran14,Bartkiewicz15}.

We also described sequences of NMR pulses to perform various
quantum tomography methods and arbitrary gates (including single
virtual qubit rotations) with nuclear spins-3/2. This enables a
simple translation of arbitrary quantum algorithms from systems of
spins-1/2 to higher-number spins.

Finally, we express our hope that this comparative study of
various NMR tomographic methods will draw attention to the issue
of how such methods are robust against errors and, thus, to the
question about the reliability of the reconstructed density
matrices.

\begin{acknowledgments}
The authors are grateful to T.~Ota, Z.~Fojud, K.~Hashimoto,
N.~Kumada, S.~Miyashita, T.~Saku, and K.~Takashina for
discussions. A.M. acknowledges a long-term fellowship from the
Japan Society for the Promotion of Science (JSPS). A.M. was
supported by the Polish National Science Centre under Grants No.
DEC-2011/03/B/ST2/01903 and No. DEC-2011/02/A/ST2/00305. J.B. was
supported by the Palack\'y University under Project No.
IGA-P\v{r}F-2014-014. N.I. was supported by JSPS Grant-in-Aid for
Scientific Research(A) (Grant No. No. 25247068). G.Y. was
supported by a Grant-in-Aid for Scientific Research (Grant No.
24241039) from the Ministry of Education, Culture, Sports, Science
and Technology (MEXT) of Japan and by the Mitsubishi Foundation.
F.N. was partially supported by the RIKEN iTHES Project, the MURI
Center for Dynamic Magneto-Optics via the AFOSR Grant No.
FA9550-14-1-0040, the IMPACT program of JST, and a Grant-in-Aid
for Scientific Research (A).
\end{acknowledgments}

\appendix

\section{Selective rotations}

Selective rotations in quadrupolar nuclei with large spins are a
simple generalization of the standard rotations in a spin-1/2
system:
\begin{eqnarray}
{\cal X}(\theta )&\equiv&  R^{x}(\theta )=\MAT{\cos
\frac{\theta}{2} & -i   \sin \frac{\theta}{2}}{-i   \sin
\frac{\theta}{2} & \cos \frac{\theta}{2} } , \label{A1}
\\ {\cal Y}(\theta
)&\equiv&  R^{y}(\theta )=\MAT{\cos \frac{\theta}{2} & - \sin
\frac{\theta}{2}}{  \sin \frac{\theta}{2} & \cos \frac{\theta}{2}
}  , \label{A2} \\
{\cal Z}(\theta )&\equiv & R^{z}(\theta)=\MAT{ e^{-i \theta/2} &
0}{0 & e^{ i\theta/2} }.  \label{A3}
\end{eqnarray}
If a two-level rotation is $R^{(i)}(\theta)=\MAT{a\;b}{c\;d}$
(with $i=X,Y,Z$), then the corresponding selective rotation
between levels $m<n$ in a $N$-level system is given by
\begin{eqnarray}
  R^{(i)}_{mn}(\theta) &=&
  a\ket{m}\bra{m}+b\ket{m}\bra{n} +c\ket{n}\bra{m}\nonumber\\
  && +d\ket{n}\bra{n} +\sum_{k\neq  n,m}\ket{k}\bra{k}.
\label{A4}
\end{eqnarray}
For example, the matrix representation of the rotation ${\cal
X}_{02}(\halfpi)$ in a spin-3/2 system reads as:
\begin{eqnarray}
   {\cal X}_{02}(\halfpi) = \frac{1}{\sqrt{2}}
\left(
\begin{array}{cccc}
 1 & 0 &-i & 0 \\
 0 & \sqrt{2} & 0 & 0 \\
-i & 0 & 1 & 0 \\
 0 & 0 & 0 & \sqrt{2}
\end{array}
\right). \label{N35}
\end{eqnarray}
Note that the rotations calculated by $\exp(-i{\cal H}_{\rm rot}
t_p/\hbar)$ are, in general, not exactly corresponding to
Eq.~(\ref{A4}), because these depend on the quadrupolar frequency
$\omega_Q$, even if the conditions $\hbar\omega_{_{\rm RF}}^{(k)}
=\epsilon_{m}-\epsilon_{n}$ and $|\omega_k| \ll |\omega_Q| \ll
|\omega_{0}|$ are satisfied~\cite{ErnstBook}. Nevertheless, these
rotations can be effectively reduced to Eq.~(\ref{A4}) if the
pulse duration $t_p$ is equal to $2\pi/\omega_Q$ or its multiple.
To fulfill this condition experimentally, the line intensities of
spectra can be monitored as a function of the pulse duration (see,
e.g., Ref.~\cite{Bonk04}).

\section{$M_{xy}$ detection vs $M_{z}$ detection}

The $M_{xy}$ detection of a spin-3/2 system provides directly the
following off-diagonal elements (as marked in boxes) of the
corresponding density matrix $\rho$:
\begin{eqnarray}
  \rho &=& \MATT{\rho_{00} & \frame{$\rho_{01}$} & \rho_{02} & \rho_{03}}
                     {\frame{$\rho_{10}$} & \rho_{11} & \frame{$\rho_{12}$} & \rho_{13}}
                     {\rho_{20} & \frame{$\rho_{21}$} & \rho_{22} & \frame{$\rho_{23}$}}
                     {\rho_{30} & \rho_{31} & \frame{$\rho_{32}$} &
                     \rho_{33}}.
\label{N04}
\end{eqnarray}
This is because the NMR signals obtained by the $M_{xy}$ detection
can be proportional to~\cite{SlichterBook}
\begin{eqnarray}
  M^{\pm}_{xy} &\equiv & M_x \pm \I  M_y \;\propto\; \tr[\rho I_{\pm}],
\label{N05}
\end{eqnarray}
as given in terms of the total angular momentum operator $I_{\pm} =
I_x {\pm} i I_y$ for spin $I=3/2$, where
\begin{equation}
 I_x =\MATT
 {0  & a   & 0   & 0}
 {a  & 0   & 1   & 0}
 {0  & 1   & 0   & a}
 {0  & 0   & a   & 0},
 \quad  I_y =i \MATT
 {0 & -a &  0 &  0}
 {a &  0 & -1 &  0}
 {0 &  1 &  0 & -a}
 {0 &  0 &  a &  0},
\label{N06}
\end{equation}
with $a=\sqrt{3}/2$. In contrast to this, the $M_{z}$ detection of
a spin-3/2 system and two spin-1/2 systems enables the
determination of only the diagonal elements $\rho_{ii}$
($i=0,...,3$). This is because the NMR signals obtained by the
$M_z$ detection of a spin-3/2 system are given by
Eqs.~(\ref{OA2}), (\ref{OA2extra}), or~(\ref{OA3}).


\end{document}